\documentclass[11pt]{article}

\usepackage{amsmath,amssymb}
\usepackage{slashed}
\usepackage{xcolor} 
\usepackage{graphicx}
\usepackage{multirow} 
\usepackage{dcolumn} 
\usepackage{bm} 
\usepackage{epsfig}
\usepackage{epstopdf}
\usepackage{grffile}
\usepackage{color}
\usepackage{colordvi}
\usepackage{amsmath,amssymb}
\usepackage{rotating}
\usepackage{lscape}
\usepackage{cite}
\usepackage{float}
\usepackage{hyperref}
\usepackage{url}
\usepackage{graphicx}
\usepackage{graphics}
\usepackage{color}
\usepackage{amssymb}
\usepackage{amsmath}
\usepackage{mathtools}
\usepackage{bm}	
\usepackage{flexisym}
\usepackage{amssymb}
\usepackage{physics}
\usepackage{mathrsfs}
\usepackage{simplewick}
\usepackage{tikz}
\usepackage{bm} 
\usepackage{makecell} 
\usepackage{color}   
\usetikzlibrary{arrows,shapes}
\usetikzlibrary{trees}
\usetikzlibrary{matrix,arrows} 			
\usetikzlibrary{positioning}			
\usetikzlibrary{calc,through}			
\usetikzlibrary{decorations.pathreplacing}  
\usepackage{pgffor}						
\usetikzlibrary{decorations.pathmorphing}
\usetikzlibrary{decorations.markings}
\tikzset{
	vector/.style={decorate, decoration={snake}, draw},
	provector/.style={decorate, decoration={snake,amplitude=2.5pt}, draw},
	antivector/.style={decorate, decoration={snake,amplitude=-2.5pt}, draw},
	fermion/.style={draw=black, postaction={decorate},
		decoration={markings,mark=at position .55 with {\arrow[draw=black]{>}}}},
	fermionbar/.style={draw=black, postaction={decorate},
		decoration={markings,mark=at position .55 with {\arrow[draw=black]{<}}}},
	fermionnoarrow/.style={draw=black},
	gluon/.style={decorate, draw=black,
		decoration={coil,amplitude=4pt, segment length=5pt}},
	scalar/.style={dashed,draw=black, postaction={decorate},
		decoration={markings,mark=at position .55 with {\arrow[draw=black]{>}}}},
	scalarbar/.style={dashed,draw=black, postaction={decorate},
		decoration={markings,mark=at position .55 with {\arrow[draw=black]{<}}}},
	scalarnoarrow/.style={dashed,draw=black},
	electron/.style={draw=black, postaction={decorate},
		decoration={markings,mark=at position .55 with {\arrow[draw=black]{>}}}},
	bigvector/.style={decorate, decoration={snake,amplitude=4pt}, draw},
}
\tikzstyle{block} = [draw, rectangle, minimum height=3em, minimum width=6em]
\usepackage{hyperref}

\usepackage{titling}
\usepackage{subcaption}
\newcommand{\subtitle}[1]{%
	\posttitle{%
		\par\end{center}
	\begin{center}\large#1\end{center}
	\vskip0.5em}%
}

\def\Re{{\cal R \mskip-4mu \lower.1ex \hbox{\it e}\,}}
\def\Im{{\cal I \mskip-5mu \lower.1ex \hbox{\it m}\,}}

\def\tev{\,{\ifmmode\mathrm {TeV}\else TeV\fi}}
\def\gev{\,{\ifmmode\mathrm {GeV}\else GeV\fi}}
\def\mev{\,{\ifmmode\mathrm {MeV}\else MeV\fi}}
\def\to{\rightarrow}

\mathchardef\mhyphen="2D 

\begin{document}

\begin{center}

\vspace*{15mm}
\vspace{1cm}
{\Large \bf Exploring non-standard $Hb\bar{b}$ interactions at future electron-proton colliders}

\vspace{1cm}

{\bf Gholamhossein Haghighat$^{1}$, Reza Jafari$^{1,2}$, Hamzeh Khanpour$^{3,4,1}$ and\\ Mojtaba Mohammadi Najafabadi$^{1}$}

 \vspace*{0.5cm}

{\small\sl 
$^{1}$School of Particles and Accelerators, Institute for Research in Fundamental Sciences (IPM) \\P.O. Box 19395-5531, Tehran, Iran. } \\
{\small\sl 
$^{2}$Department of Physics, University of Tehran, North Karegar Avenue, Tehran 14395-547, Iran. } \\
{\small\sl 
$^{3}$AGH University, Faculty of Physics and Applied Computer Science, \\Al. Mickiewicza 30, 30-055 Kraków, Poland.} \\
{\small\sl 
$^{4}$Department of Physics, University of Science and Technology of Mazandaran, \\P.O.Box 48518-78195, Behshahr, Iran.} \\
\vspace*{.2cm}
\end{center}

\vspace*{10mm}

%
%
\begin{abstract}\label{abstract}

In this paper, we use the charged-current Higgs boson production 
process at future electron-proton colliders, $e^-p \to H j \nu_e$, with the subsequent decay of the Higgs boson into a $b\bar{b}$ pair,
to probe the Standard Model effective field theory
with dimension-six operators involving the Higgs boson and the bottom quark.
The study is performed for two proposed future high-energy electron-proton colliders, 
the Large Hadron Electron Collider (LHeC) and the 
Future Circular Collider (FCC-he) at 
the center-of-mass energies of 1.3 TeV and 3.46 TeV, respectively.
Constraints on the CP-even and CP-odd $Hb\bar{b}$ couplings are derived by analyzing the simulated signal and background samples. A realistic detector simulation is performed and a multivariate technique using the gradient Boosted Decision Trees algorithm is employed to discriminate the signal from background. Expected limits are obtained at $95\%$ Confidence Level for the LHeC and FCC-he assuming the integrated luminosities of 1, 2 and 10 ab$^{-1}$.
We find that using 1 ab$^{-1}$ of data, the CP-even and CP-odd $Hb\bar{b}$ couplings can be constrained with accuracies of the order of $10^{-3}$ and $10^{-2}$, respectively, and a significant region of the unprobed parameter space becomes accessible.
\end{abstract}
\vspace{2cm}

\newpage

%
\section{Introduction}\label{sec:intro}

The Higgs boson, the latest discovered elementary particle in the Standard Model (SM), was observed at the Large Hadron Collider (LHC) by the CMS and ATLAS experiments at CERN~\cite{Chatrchyan:2012xdj,Aad:2012tfa} 
and attracted significant attention from the particle physics community.
In recent years, many experimental and 
phenomenological studies have been performed to 
study and measure the Higgs boson properties as accurately as 
possible to test the validation of the SM and 
find deviations from the SM in order to explore any 
signatures of New Physics (NP).
In addition, the precision measurement of the Higgs boson properties, 
in particular, its coupling to other 
SM particles, could provide in-depth 
information on the electroweak symmetry 
breaking (EWSB) mechanism.
As the Higgs boson studies at the LHC has not found any clear evidence of NP yet, accurate measurement of different couplings in the Higgs sector can be one of the primary goals of the
future high energy colliders \cite{r1}.
To quantify the agreement between the signal yields in data and the expectations from the SM, 
the so-called signal strength parameter, $\mu$, is used. It scales the observed yields with respect to 
the yields predicted by the SM without any modification in the shape of the distributions. 
The signal strength for the Higgs boson decay into a pair of bottom quarks, $H\rightarrow b\bar{b}$, is defined as
 $\mu_{b\bar{b}} = \mathcal{B}_{\rm data}(H\rightarrow b\bar{b})/ \mathcal{B}_{\rm SM}(H\rightarrow b\bar{b})$, where $\mathcal{B}_{\rm data(SM)}$ denotes the observed (SM predicted) decay branching fraction.
The recent CMS and ATLAS measurements found the values \cite{CMS:2022dwd, ATLAS:2020fcp}:
\begin{eqnarray}
\text{CMS:}&\mu_{b\bar{b}} = 1.05\pm 0.15(\text{stat.})^{+0.16}_{-0.15} (\text{syst.}), \,\,\,\,\,\,& \nonumber \\
\text{ATLAS:}&\mu_{b\bar{b}} = 1.02\pm 0.12(\text{stat.})\pm 0.14(\text{syst.}),& \nonumber
\end{eqnarray}
for the Higgs boson decay signal strength. 
Both measurements are based on the LHC full Run II data recorded at the center-of-mass energy of 13 TeV. As can be seen, the current uncertainty on the measurement of the branching fraction of $H\rightarrow b\bar{b}$ decay is around $20\%$ and the observation is compatible with the prediction of the SM. The sensitivity of various future colliders to potential deformations of the Higgs boson coupling to $b\bar{b}$ with respect to the SM has been studied through the Higgs effective coupling $g^{\rm eff~2}_{Hbb} = \Gamma(H\rightarrow b\bar{b})/\Gamma_{\rm SM}(H\rightarrow b\bar{b})$.
The expected accuracy at $68\%$ Confidence Level (CL) on $g^{\rm eff}_{Hbb}$ from the global fit to the Higgs boson, di-boson and Electro-Weak Precision Observables (EWPO) at the HL-LHC is found to be $5\%$ \cite{hfull}.
The uncertainty on the Higgs boson decay measurements allows for the existence of a wide range of models beyond the SM which can be efficiently explored with the help of the Effective Field Theory (EFT) approach.

It is well-known that the SM is a low-energy effective theory which is an approximation of a more complete theory.
The most general effective Lagrangian which includes dimension-four and -six operators describing the Higgs field coupling to the third generation fermions is given by \cite{Fuchs:2020uoc}:
\begin{eqnarray}
\mathcal{L}_{\rm eff} = y_{f}\bar{F}_{L}F_{R}H+\frac{X_{R}^{f}+i X_{I}^{f}}{\Lambda^{2}}|H|^{2}\bar{F}_{L}F_{R}H + h.c., 
\label{eq:mainLagrangian}
\end{eqnarray}
where $F_{L}$ is the $\rm SU(2)$ doublet of the third generation fermions, $F_{R}$ is the corresponding $\rm SU(2)$
singlet, $H$ denotes the Higgs doublet  field, $\Lambda$ is the mass scale of NP, $y_{f}$ is the Yukawa coupling constant for fermion $f$, and $X_{R}^{f}$ and
$X_{I}^{f}$ are the real and imaginary parts of the Yukawa coupling constant originating from dimension-six operators.\footnote{{A more general effective Lagrangian may also include dimension-eight operators. Dimension-eight operators are not considered in this study. However, their effects on the process studied here may be non-negligible compared with the effect of dimension-six operators in some cases. This imposes a restriction on the validity of the results obtained in this study. We will discuss this later.}}
{Addition of this dimension-six operator to the SM Lagrangian leads to the modification of the Higgs boson coupling to fermions and has important consequences. The presence of both the CP-even and CP-odd processes in the dimension-six operator leads to novel CP-violating interactions which may open a window to solving the long-standing problem of the observed baryon asymmetry. The CP-violating interactions may contribute to the baryon asymmetry via electroweak baryogenesis (EWBG) \cite{Shu:2013uua,Grojean:2004xa,Chung:2012vg}. Furthermore, the modification of the Higgs boson couplings leads to the deviation of the Higgs boson production and decay rates from the SM predicted values. Such deviations are predicted by many models beyond the SM. In the Minimal Supersymmetric Standard Model (MSSM) \cite{Gunion:1989we,Djouadi:2005gj}, the coupling of the SM-like Higgs to fermions depends not only on the fermion mass but also on other model parameters making the SM-like Higgs production and decay rates deviate from the SM predictions. This is also the case in different types of the Two-Higgs-Doublet Model (2HDM) \cite{Branco:2011iw} where the SM-like Higgs couplings to fermions and gauge bosons are modified. Models with Axion-Like Particles (ALPs) \cite{DiLuzio:2020wdo,Sakurai:2021ipp} may also modify the Higgs boson couplings and may be probed by searching for anomalous Higgs boson couplings.}
With the use of the effective Lagrangian, Eq. \ref{eq:mainLagrangian}, the effective Lagrangian involving a Higgs boson and a pair of bottom quarks after the EWSB can be written as:
\begin{eqnarray}
\label{eq:Lagrangian}
\mathcal{L}_{\rm eff} \supset  \frac{y_{b}}{\sqrt{2}}\left[1+  3(T_{R}+i T_{I})\right] \bar{b}_{L} b_{R} h,
\end{eqnarray}
where
\begin{eqnarray}
T_{R} = \frac{ v^{2}}{2 \Lambda^{2}}\frac{X_{R}^{b}}{y_{b}} ~,~T_{I} = \frac{v^{2}}{2 \Lambda^{2}}\frac{X_{I}^{b}}{y_{b}},
\end{eqnarray}
with $v$ being the vacuum expectation value of the Higgs field. It is notable that the effective Lagrangian, Eq. \ref{eq:mainLagrangian}, also leads to couplings involving two and three Higgs bosons. Such couplings are not relevant to the present study and thus are not shown in Eq. \ref{eq:Lagrangian} (see Ref. \cite{Fuchs:2020uoc} for the explicit form of such couplings). The CP-odd part of the effective Lagrangian, Eq. \ref{eq:Lagrangian}, with the coupling constant $T_{I}$ originating from the dimension-six operators contributes to the electric dipole moment (EDM) of the electron, $d_{e}$, through loops. 
As a result, the magnitude of $T_{I}$ can be indirectly constrained using the current experimental bound on $d_{e}$. 
The finite contribution arising from the Barr-Zee diagrams due to $T_I$ has the following form:
\begin{eqnarray}
|d_{e}| =  N_{c}Q_{b}^{2}\times\frac{\sqrt{2}\alpha}{\pi^3}\frac{m_{e}}{m_{h}^{3}}\frac{m_{b}^{2}}{v^{2}} |T_{I}| \left[ \frac{\pi^2}{3} + \ln^{2}\left(\frac{m_{b}^{2}}{m_{h}^{2}}\right) \right],
\end{eqnarray}
where $N_{c}$ is the number of colors and $Q_{b} = -1/3$ is the bottom quark electric charge.
The experimental bound on the electron EDM at $90\%$ CL derived by the ACME collaboration is $|d_e| \leq 1.1\times 10^{-29}$ $\rm e\cdot cm$ \cite{ACME:2018yjb}. 
Based on the ACME bound on the electron EDM, the bound on $T_{I}$ at $90\%$ CL is found to be \cite{Fuchs:2020uoc}: 
\begin{eqnarray}
T_{I} \in [-0.09 , 0.09].
\end{eqnarray}
The limit derived from the electron EDM measurements is valid under the assumption that the $He\bar{e}$ coupling has no deviation from the SM and no cancellations with other contributing mechanisms exist.
Direct searches through the measurement of the Higgs boson signal strength $\mu_{b\bar{b}}$ lead to the following constraints on $T_{I}$ and $T_{R}$ \cite{Fuchs:2020uoc}:
\begin{eqnarray}
T_{I} \in [-0.43 , 0.43] ~,~  T_{R} \in [-0.55,-0.46]\cup[-0.12,0.33].
\end{eqnarray}
Hence, the most stringent experimental bounds on $T_{I}$ and $T_{R}$ are currently derived from the measurements of the electron EDM and the Higgs boson signal strength, respectively.

This study is dedicated to constraining the $T_{I}$ and $T_{R}$ coupling constants at future electron-proton colliders to investigate the capability of these colliders to probe the $Hb\bar{b}$ dimension-six couplings. The future colliders LHeC \cite{LHeCStudyGroup:2012zhm} and FCC-he \cite{LHeC:2020van} with the respective planned operating energies 1.3 and 3.46 TeV are assumed in this study. In an electron-proton collider, the center-of-mass energy is given by $\sqrt{s} = 2 \sqrt{E_eE_p}$, where $E_e$ and $E_p$ denote the energies of the electron and proton beams, respectively.
Based on the present proposals, the proton beam energy is assumed to be 7 and 50 TeV at the LHeC and FCC-he, respectively, and the electron beam energy is assumed to be 60 GeV for both of them \cite{LHeCStudyGroup:2012zhm,LHeC:2020van}.
There are two main Higgs boson production mechanisms at electron-proton colliders:
the so-called neutral-current (NC) production, $e^{-}q \rightarrow e^{-}qH$, and the so-called charged-current (CC) production, $e^{-}q \rightarrow \nu_{e}q^\prime H$. 
The production rate for the CC channel is larger than the NC channel by a 
factor of around 5 \cite{Han:2009pe}. 
In this work, the effective $Hb\bar{b}$ coupling is studied using the CC channel Higgs boson production followed by the Higgs boson decay into a pair of bottom quarks. 
Different aspects of the Higgs boson physics at the LHC, LHeC and FCC-he have been extensively studied to date. Refs. \cite{Behera:2022gnr, Han:2009pe, Biswal:2012mp,Senol:2012fc, Hesari:2018ssq, Dutta:2021del,  FCC:2018byv, Blumlein:1992eh, LHeC:2020van, Khatibi:2014bsa, Han:2023krp} include a number of such studies on observability of the SM Higgs boson,
prospects for the measurement of the Yukawa coupling constants of the third generation fermions, and
the Standard Model Effective Field Theory (SMEFT).

This paper is organized as follows.
In section~\ref{sec:Theoretical}, the CC channel Higgs boson production followed by the $H\rightarrow b\bar{b}$ decay at the LHeC and FCC-he is studied considering the $Hb\bar{b}$ effective couplings.  
In section \ref{sec:Simulation}, the Monte Carlo 
(MC) event generation, the detector simulation, and the analysis strategy in this study are described.
Section \ref{sec:Results} is devoted to the results and the discussion on our findings. Finally, section~\ref{sec:Conclusion} concludes and summarizes the paper.

%
\section{Charged current Higgs boson production}
\label{sec:Theoretical}

\begin{figure*}[htb]
\begin{center}
\resizebox{0.215\textwidth}{!}{\includegraphics{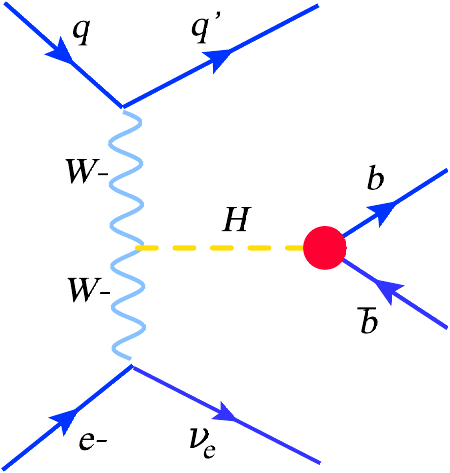}}
\caption{{\small Leading-order Feynman diagram for the Higgs boson
production via the CC channel at an electron-proton collider, 
where the Higgs boson decays into a $b \bar{b}$ pair. 
The red filled circle shows the vertex affected by the dimension-six operators.} 
\label{Fig:signal_diagram}}
\end{center}
\end{figure*}

The main Higgs boson production processes in the electron-proton collisions, the CC and NC production channels, proceed through the vector boson fusion \cite{Han:2009pe}. The leading-order Feynman diagram for the Higgs boson production through the CC channel with subsequent decay of the Higgs boson into a pair of bottom quarks, $e^{-}p \rightarrow Hj\nu_{e} \rightarrow b\bar{b}+j\nu_{e}$ ($j=u,d,c,s,b$), is shown in Fig. \ref{Fig:signal_diagram}.
As mentioned earlier, in this work, the focus is on the Higgs boson 
production through the CC channel because it has a larger production cross section than the NC channel.
It can be shown that the cross section of the Higgs boson production through the CC channel is larger than the production cross section corresponding to the NC channel by a factor of $\sim 5.3$ at the LHeC ($E_{e} = 60$ GeV and $E_{p} = 7$ TeV) and  by a factor of $\sim 4.6 $ at the FCC-he ($E_{e} = 60$ GeV and $E_{p} = 50$ TeV). 

\begin{figure*}[htb]
\vspace{0.50cm}	
\begin{center}
\includegraphics[width=0.49\textwidth]{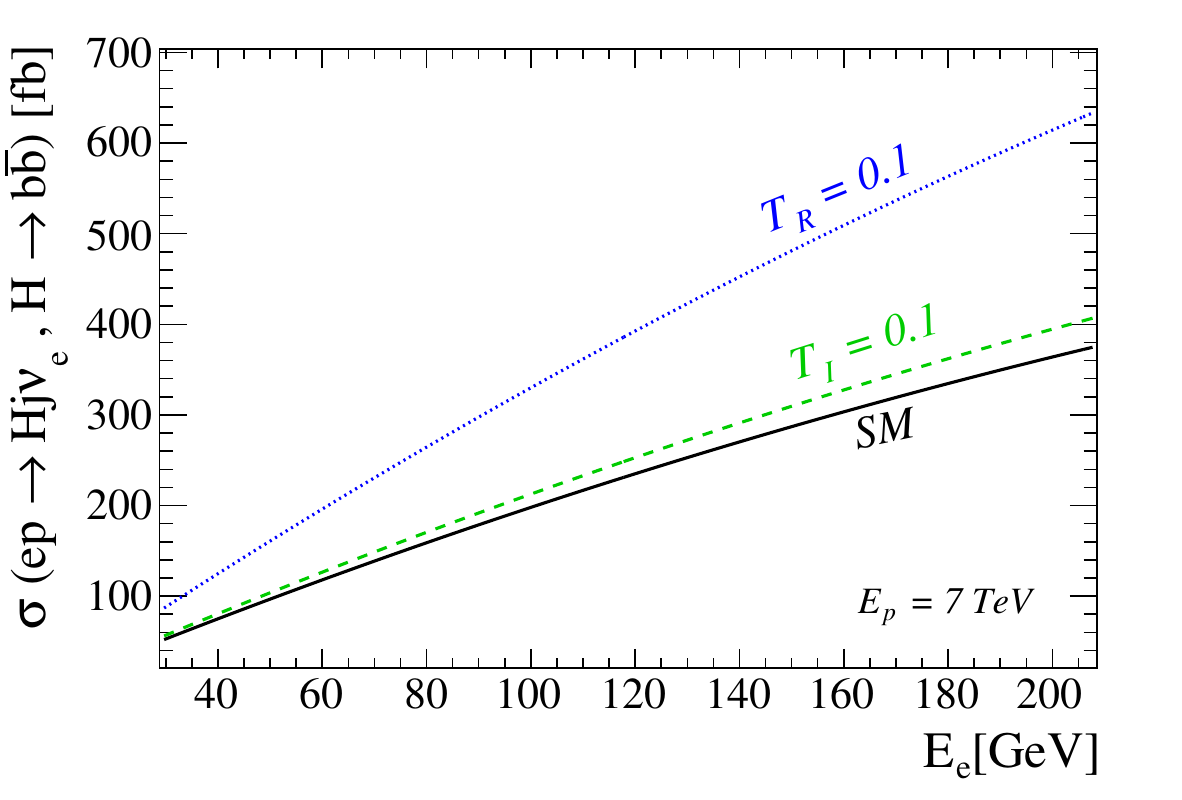}
\includegraphics[width=0.49\textwidth]{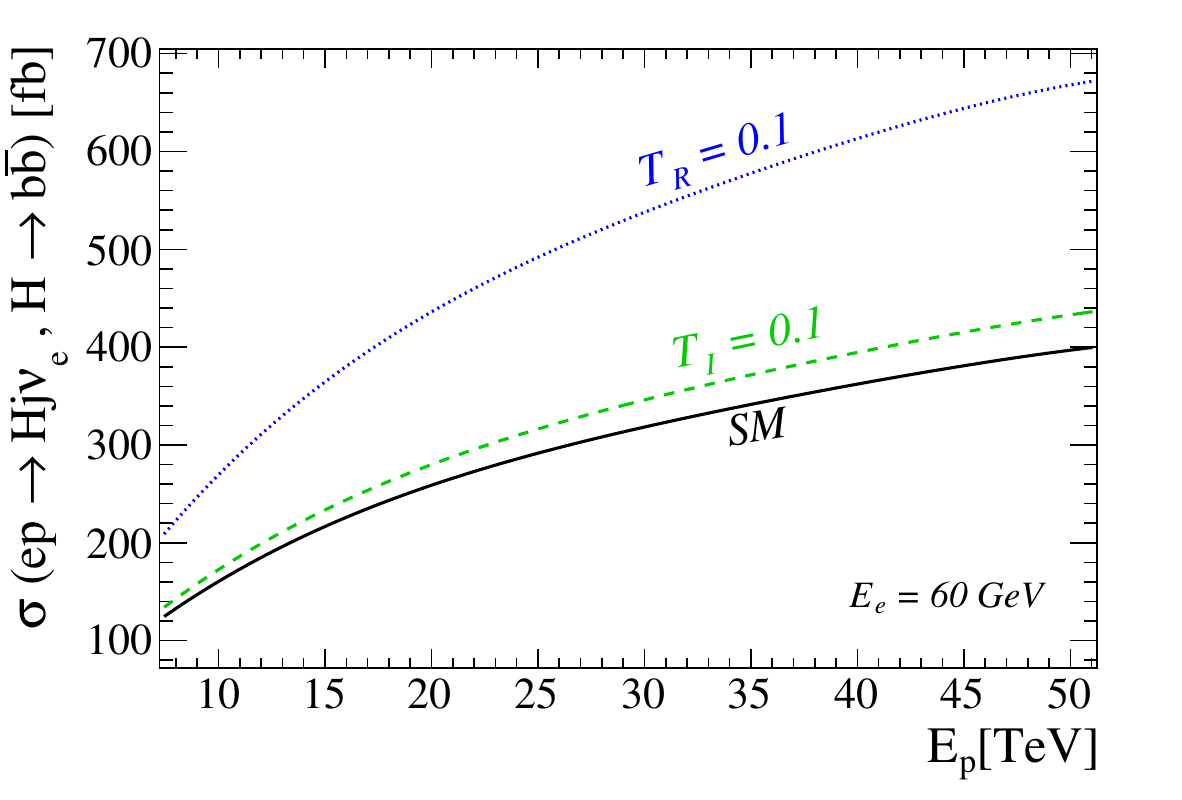}
\caption{{\small  
Cross section of $H j \nu_e$ production at the leading-order through the CC channel in $e^-p$ collisions with the subsequent $H \to b\bar{b}$ decay as a function of the electron beam energy where the energy of the proton beam is set to 7 TeV (left), and as a function of the proton beam energy where the energy of the electron beam is set to 60 GeV (right). Cross sections are provided for the three cases: 1) no dimension-six coupling is present (black solid line), 2) $T_I = 0.1$, $T_R = 0$ (green dashed line) and 3) $T_R = 0.1$, $T_I = 0$ (blue dotted line).
} \label{fig:sigma_Ee}}
\end{center}
\end{figure*}

There are different proposals with different energy scenarios for possible future electron-proton colliders. The energy of the electron beam in these proposals ranges from 50 GeV to 200 GeV. 
Fig. \ref{fig:sigma_Ee} shows the cross section of the leading-order CC channel Higgs boson production in $e^-p$ collisions followed by the $H\rightarrow b\bar{b}$ decay with and without the presence of the dimension-six $Hb\bar{b}$ couplings versus the energies of the electron and proton beams.
To compute the cross sections, the proton parton distribution functions (PDFs) are taken from {\tt NNPDF2.3} \cite{Ball:2012cx,NNPDF:2017mvq}, and {\tt MadGraph5\_aMC@NLO}~\cite{Alwall:2011uj,Alwall:2014bza,Alwall:2014hca} is used for the matrix element calculations. To take the effects of the dimension-six operators into account, the effective Lagrangian, Eq. \ref{eq:Lagrangian}, is implemented into FeynRules \cite{Degrande:2011ua} and the generated Universal FeynRules Output (UFO) model is passed to {\tt MadGraph5\_aMC@NLO}.
\begin{figure*}[!htb]
  \centering  
    \begin{subfigure}[b]{0.49\textwidth}
    \centering
    \includegraphics[width=\textwidth]{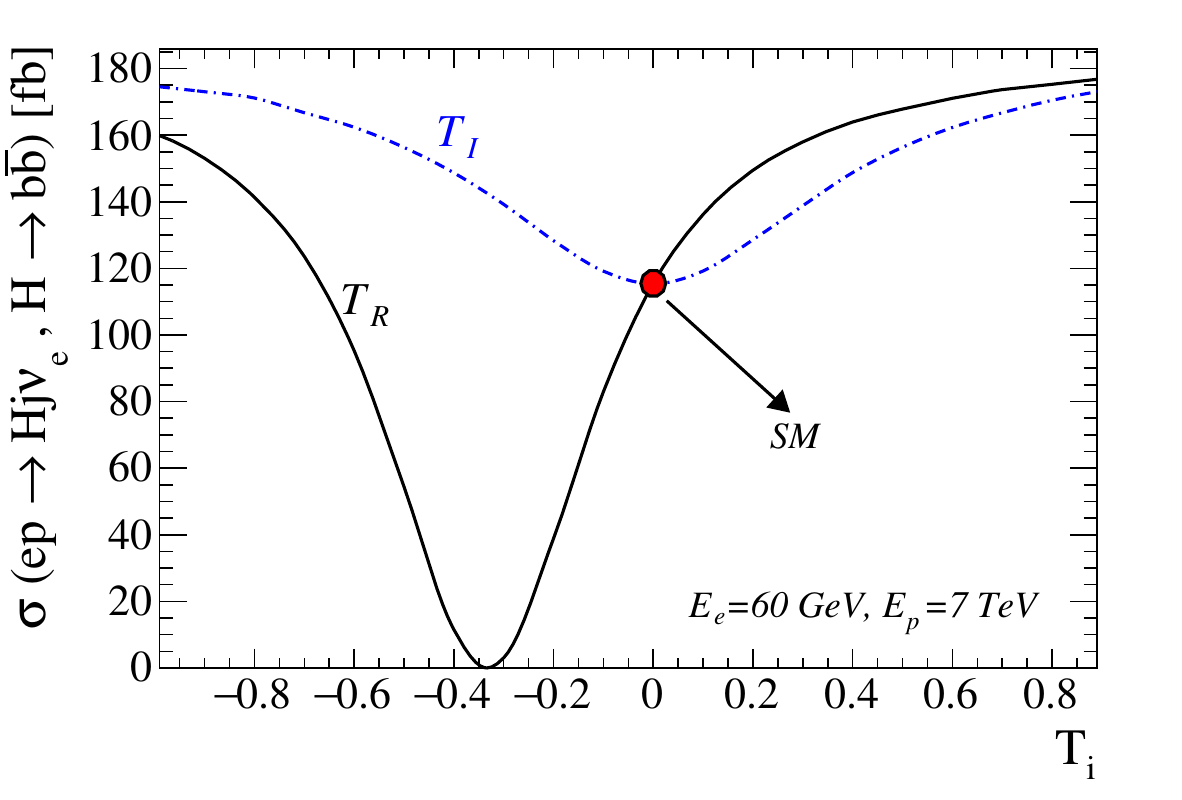}
    \caption{}
    \label{sigma_couplings}
    \end{subfigure} 
    \begin{subfigure}[b]{0.49\textwidth} 
    \centering
    \includegraphics[width=\textwidth]{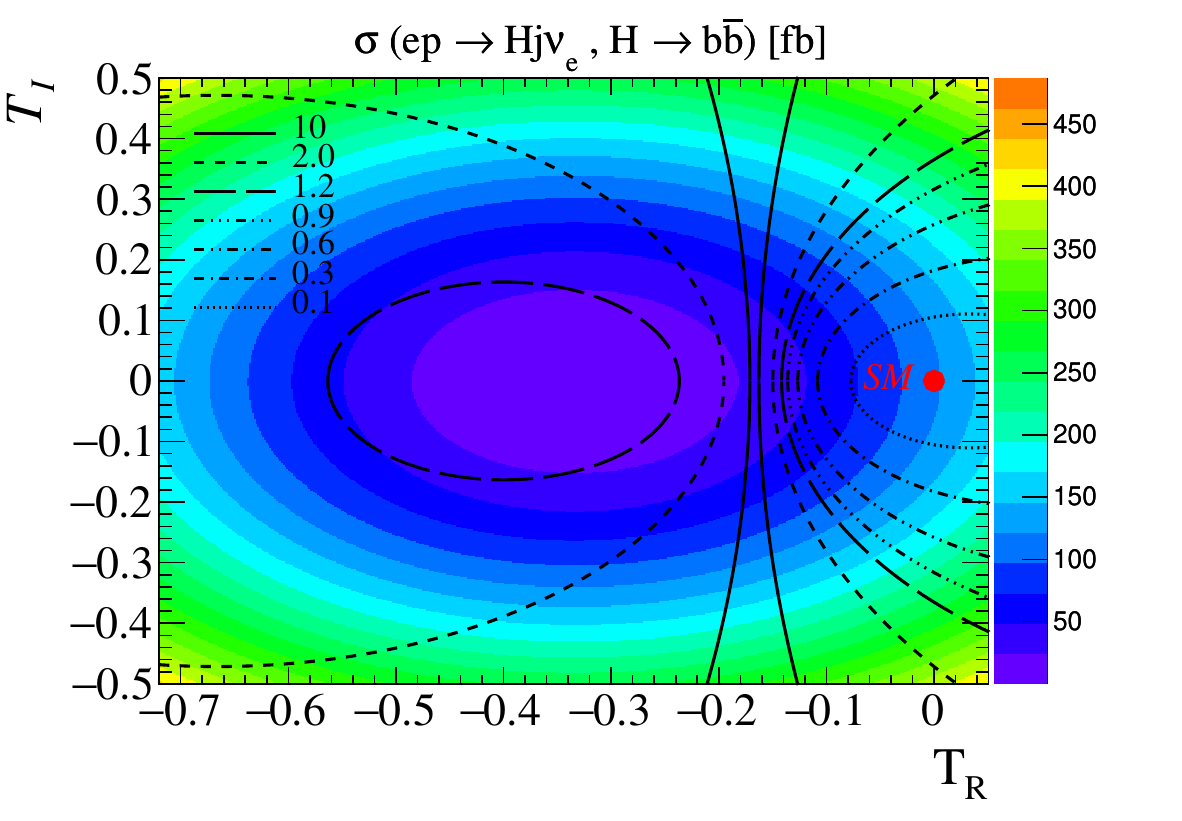}
    \caption{}
    \label{xsec2dim_plot}
    \end{subfigure}
\caption{{\small 
{Cross section of the CC process $e^{-}p\rightarrow H j \nu_e, \, H \rightarrow b\bar{b}$ at the leading-order a) assuming one non-zero coupling constant at a time and b) as a function of the $T_{R}$ and $T_{I}$ coupling constants at a center-of-mass energy of $\sqrt{s}=1.3$ TeV ($E_{e} = 60$ GeV, $E_{p} = 7$ TeV). The contour lines on the right-hand side plot show the value of the ratio $\mathcal{R}$ at a given point of the parameter space.}} 
\label{fig:sigma_X}}
\end{figure*}
The left plot corresponds to the proton beam energy of 7 TeV and the right plot corresponds to the electron beam energy of 60 GeV. 
The cross section can be parameterized as:
\begin{eqnarray}
\sigma(e^{-}p \rightarrow Hj\nu_{e}, \, H \rightarrow b\bar{b}) (T_R,T_I) \approx \sigma_{\rm SM} +  701\, T_R + 1042\, T_R^{2} + 1052\, T_I^{2} \,\,\, \mathrm{fb},
\label{xsecpara}
\end{eqnarray}
where $\sigma_{\rm SM} \approx 116.9$ fb is the cross section of the same process when no dimension-six operator is present. As the total cross section is a CP-even observable, the cross section is not linearly dependent on $T_{I}$. {The cross section, Eq. \ref{xsecpara}, contains a term of order $1/\Lambda^2$ as well as terms of order $1/\Lambda^4$. If one includes dimension-eight operators in the effective Lagrangian, Eq. \ref{eq:mainLagrangian}, corrections of orders $1/\Lambda^4$ and $1/\Lambda^8$, stemming from dimension-eight operators, will also be added to the corrections already included in Eq. \ref{xsecpara}. It can be seen that, in some region of the parameter space, the effect of the $1/\Lambda^4$ corrections present in Eq. \ref{xsecpara} is small compared with the contribution of terms up to order $1/\Lambda^2$. However, there are regions of parameter space where the contribution of the $1/\Lambda^4$ terms is not small and negligible. In such regions, dimension-eight operators may also induce important effects through $1/\Lambda^4$ corrections and cannot be neglected. This study only considers dimension-six operators. Therefore, the results obtained in this study are only valid for regions of parameter space where the contribution of $1/\Lambda^4$ corrections is small. In other regions, the effect of dimension-eight operators should also be taken into account for the EFT approach to be valid. Fig.~\ref{fig:sigma_X} shows the cross section of the same process at a center-of-mass energy of $\sqrt{s}=1.3$ TeV. Fig. \ref{sigma_couplings} assumes that only one of the coupling constants is non-zero and Fig. \ref{xsec2dim_plot} shows the cross section as a function of the $T_R$ and $T_{I}$ coupling constants through a heatmap.} The cross section is computed using the same method as in Fig. \ref{fig:sigma_Ee}. As seen, the cross section is strongly dependent on $T_{R}$. This is because $T_{R}$ contributes to both the $1/\Lambda^2$ and $1/\Lambda^4$ corrections while $T_{I}$ only contributes to the $1/\Lambda^4$ correction (see Eq. \ref{xsecpara}). The $1/\Lambda^2$ correction, which stems from the interference term involving $T_{R}$, leads to an asymmetric shape for the cross section. This asymmetry leads to asymmetric constraints on $T_{R}$. {It can be seen that the cross section vanishes at $T_R=-1/3$, $T_I=0$ and recovers its SM value at any point on the circumference of a circle with a radius of 1/3 centered at $T_R=-1/3$, $T_I=0$ (see Fig. \ref{xsec2dim_plot}). As the main focus is on regions of parameter space with small $1/\Lambda^4$ corrections, the ratio of the contribution of the $1/\Lambda^4$ terms to the contribution of the remaining terms in Eq. \ref{xsecpara}, $\mathcal{R} \equiv (1042\, T_R^{2} + 1052\, T_I^{2})/|\sigma_{\rm SM} +  701\, T_R| $, is calculated and the results are shown by the contour lines in Fig. \ref{xsec2dim_plot}. The dotted, dash-dotted and dash-double-dotted lines respectively correspond to the $\mathcal{R}$ values 0.1, 0.3 and 0.6. Other contour lines correspond to higher $\mathcal{R}$ values. The ratio $\mathcal{R}$ becomes infinitely large in the vicinity of the line $T_R=-1/6$, where the contribution of the interference term ($701\,T_R$) cancels out the SM cross section. At any point of the parameter space on the left side of this line, $\mathcal{R}$ is found to be $\gtrsim1$. It is seen that the parameter space regions with small $1/\Lambda^4$ corrections are the regions around the point $T_R=T_I=0$. In regions far from the zero point, the $1/\Lambda^4$ corrections are non-negligible and thus dimension-eight operators should also be considered as they may have significant effects.}

{
Validity of the effective Lagrangian, Eq. 1, can be ensured by requiring the mass scale of new physics, $\Lambda$, to be significantly larger than the typical energy scale of the process under study. All the events should be therefore required to satisfy the condition $\sqrt{\hat{s}}<\Lambda$ to strictly ensure the validity of the EFT. The experimental measurement of $\sqrt{\hat{s}}$ is, however, only possible if all the particles in the process final state are detectable, which is not the case in this study due to the presence of undetectable neutrinos. Considering the correlation between $\sqrt{\hat{s}}$ and the missing transverse energy, obtained using the Monte Carlo simulated events, validity of the EFT may be naively ensured by requiring $2\slashed{E}_T^{max}<\Lambda$, where $\slashed{E}_T^{max}$ denotes the highest missing transverse energy data bin in the analysis. In this study, the strict EFT validity condition $\sqrt{\hat{s}}<\Lambda$ is imposed by discarding the fraction of events for which $\sqrt{\hat{s}}>2\slashed{E}_T^{max}$, where $\slashed{E}_T^{max}$ is obtained to be around 0.5 and 1 TeV at the LHeC and FCC-he, respectively.
}

\section{Monte Carlo simulation and the analysis strategy}\label{sec:Simulation}

\begin{figure*}[t]
\begin{center}
\resizebox{0.82\textwidth}{!}{\includegraphics{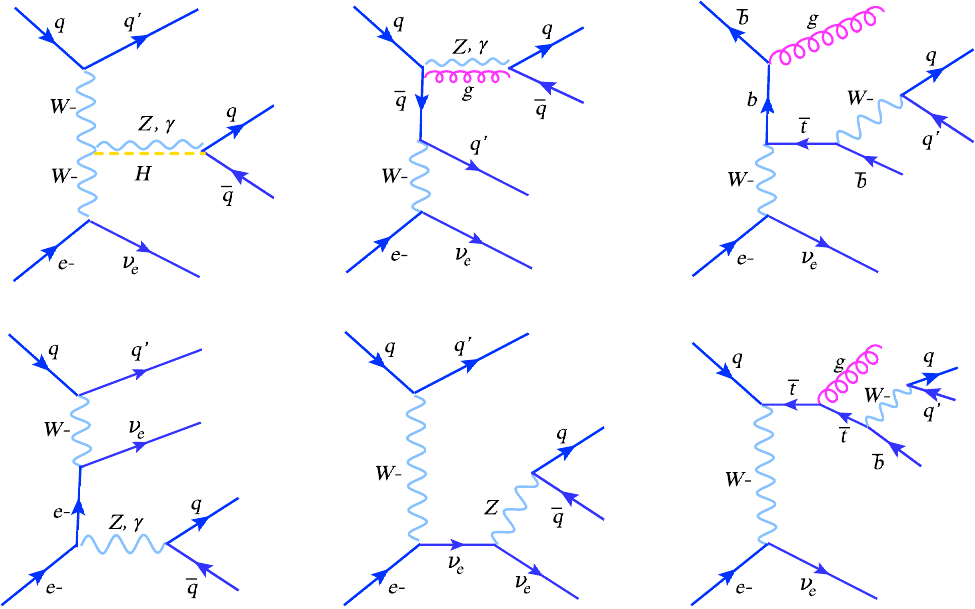}}
\caption{{\small 
Representative Feynman diagrams at the leading-order for the dominant SM background processes.
} \label{Fig:bkg_diagrams}}
\end{center}
\end{figure*}

The signal final state includes three jets and a large missing transverse energy mainly due to the missing electron neutrino and neutrinos produced in the decay of unstable hadrons inside jets. Out of the three jets in the final state, two jets are $b$-flavored and one jet can have any flavor (this jet is likely to be in the forward region because of the nature of $t$-channel processes). The main SM background processes relevant to the signal process are as follows: 
\begin{itemize}
\item  $e^-p \to H  j \nu_e,\, H \to b\bar{b}$ (SM irreducible background), 
\item $e^-p \to Z j \nu_e, \, Z \to b\bar{b},$ 
\item $e^-p \to Z  j \nu_e, \, Z \to c\bar{c},$  
\item  $e^-p  \to V^{*} j \nu_e, \, V^{*} \to b\bar{b} \,\, (V = \gamma,g),$ 
\item  $e^-p  \to V^{*} j \nu_e, \, V^{*} \to c\bar{c} \,\, (V = \gamma,g),$ 
\item  $e^-p  \to V^{*} j \nu_e, \, V^{*} \to j_{\ell}\bar{j}_{\ell} \,\, (V = Z,\gamma,g, \, j_{\ell}=u,d,s),$  
\item  $e^-p \to \bar{t} j \nu_e, \, \bar{t}\to W^- \bar{b}, \, W^- \to jj.$ 
\end{itemize}
The $H(\to b\bar{b}) j \nu_e$ background process is irreducible. This process is the SM counterpart of the signal process and cannot be suppressed. The rest of the background processes produce the same signature as the signal if the light $uds$ jets or charm-quark jets are misidentified as the $b$-jet or if a $b\bar{b}$ pair is produced through the conversion of a gauge boson. The background processes listed above are considered in this study and their contributions are estimated. It should be noted that the processes $e^-p \to c\bar{c} j \nu_e$ and $e^-p \to  j_{\ell}\bar{j}_{\ell} j \nu_e$ can also proceed through one or two intermediate $W$ bosons at the tree-level. These subdominant processes have also been included in the processes $e^-p \to V^*(\to c\bar{c}) j \nu_e$ and $e^-p \to V^*(\to  j_{\ell}\bar{j}_{\ell}) j \nu_e$ in the above list (5th and 6th processes) although it is not stated explicitly. In addition to the above-mentioned processes, there are also a number of subdominant background processes with small contributions to the total background. The photo-production of a pair of bottom quarks in association with a jet, $e^{-}p \to e^{-} p \gamma \to e^{-}j b\bar{b}$, is an example of such processes. The contribution of this process to the total background is found to be less than $1\%$. The effects of such processes can be safely neglected compared with the systematic uncertainty considered in section \ref{sec:Results}. The representative Feynman diagrams at the leading-order for the main SM background processes are shown in Fig.~\ref{Fig:bkg_diagrams}.

The signal and background processes are simulated and analyzed assuming the center-of-mass energies of 1.3 and 3.46 TeV according 
to the LHeC and FCC-he planned energy scenarios. The simulation and analysis for the two energy scenarios are performed independently. FeynRules ~\cite{Degrande:2011ua} package is employed for the implementation of the effective Lagrangian, Eq. \ref{eq:Lagrangian}. The UFO model generated by FeynRules is passed to {\tt MadGraph5\_aMC@NLO}~\cite{Alwall:2011uj,Alwall:2014bza,Alwall:2014hca} to generate hard events.
Parton showering, hadronization and decay of unstable particles are performed using {\tt PYTHIA 8.3}~\cite{Sjostrand:2014zea,Sjostrand:2007gs}. The produced events are internally passed to {\tt Delphes 3.4.2}~\cite{deFavereau:2013fsa} to simulate the detector effects using the FCC-he~\cite{gitFCCeh:2020} and LHeC detector cards~\cite{gitLHeC:2020}. Jets are reconstructed with the use of the anti-$k_{t}$ algorithm~\cite{Cacciari:2008gp} inside the {\tt FastJet 3.3.2} package~\cite{Cacciari:2011ma} assuming the cone size of 0.4 (AK4). The $b$-tagging efficiency is assumed to be 75\% and the misidentification rate to tag a $c$-jet (light-jet) as a $b$-jet is set to $5\%$ ($0.1\%$) for the LHeC detector. 
For the FCC-he detector, the $b$-tagging efficiency and the misidentification rates are functions of the transverse momentum and pseudorapidity of the candidate jet. The maximum $b$-tagging efficiency for the FCC-he detector is assumed to be 85$\%$ and is applied to jets with $|\eta|<2.5$ and $p_T<500$ GeV. Furthermore, the $c$-jet and light-jet misidentification rates are respectively 4$\%$ and 0.1$\%$ at most.

The event selection performed in this analysis is as follows. Events are required to have exactly three AK4 jets with transverse momenta greater than 30 GeV. Two out of the three jets should be tagged as $b$-jets and the third jet should not be $b$-tagged. The two $b$-tagged jets in the signal process are produced in the decay of the Higgs boson and can be used to reconstruct the Higgs boson. The pseudorapidity of the $b$-jets should satisfy the condition $|\eta_{b\mhyphen jet}| \leq 2.5$ and the pseudorapidity of the third (forward) jet is required to be in the range $|\eta_{jet}| \leq 5$. Jets are required to be well-separated by imposing the criterion $\Delta R>0.5$ on any two jets in the event. The distance $\Delta R$ is defined as $\Delta R = \sqrt{\Delta\eta^2+\Delta\phi^2}$, where $\eta$ and $\phi$ are respectively the pseudorapidity and azimuth angle. Events are required to contain no isolated electrons and muons. Isolated objects are identified using the relative isolation variable $I_{rel}$, defined as $I_{rel}=\Sigma \, p_T^{\,i}/p_T^{\,\mathrm{P}}$, where P is the candidate particle and $i$ is the summation index running over all particles (excluding the particle P) within a cone centered on the candidate particle. The cone has a size of 0.4 (0.3) for the LHeC (FCC-he). An electron or a muon is identified as an isolated object if $I_{rel}<0.1$. Isolated electrons and muons should satisfy the conditions $p_T>10$ GeV and $\vert \eta \vert < 2.5$. Finally, the missing transverse energy, $\slashed{E}_T$, is required to be greater than 20 GeV in all events. Applying all the above-mentioned event selection cuts, the selection efficiencies provided in Tab. \ref{table:eff} are obtained for the signal and background processes. As seen, the SM background processes $H(\to b\bar{b}) j \nu_e$, $Z(\to b\bar{b}) j \nu_e$, ${\gamma/g}^{*}(\to b\bar{b}) j \nu_e$ and $\bar{t} j \nu_e$ have higher event selection efficiencies as expected. Processes with a $c\bar{c}$ or $ j_{\ell}\bar{j_{\ell}}$ pair in the final state have very low event selection efficiencies due to the smallness of the misidentification probability to tag a $c$-jet or light-jet as a $b$-jet.

Preselected events are analyzed using a set of well-chosen discriminating variables to separate the signal from background. The discriminating variables used in this analysis are listed below. Note that the variables are assumed to be measured in the laboratory frame unless stated otherwise.
\begin{table}[t]
	\footnotesize
	\begin{center}
		\begin{tabular}{ 
				>{\centering\arraybackslash}m{.6in} 
				|>{\centering\arraybackslash}m{.9in}  
				>{\centering\arraybackslash}m{.9in}  
				>{\centering\arraybackslash}m{1.1in} 
				>{\centering\arraybackslash}m{.7in}
				>{\centering\arraybackslash}m{0in}} 
			\parbox{0pt}{\rule{0pt}{.8ex+\baselineskip}}{$\sqrt s$ [TeV]} & 
			\parbox{0pt}{\rule{0pt}{.8ex+\baselineskip}}{$T_R=0.1$} & 
			\parbox{0pt}{\rule{0pt}{.8ex+\baselineskip}}{$H(\to b\bar{b}) j \nu_e$} & \parbox{0pt}{\rule{0pt}{.8ex+\baselineskip}}{$Z(\to b\bar{b}) j \nu_e$} & \parbox{0pt}{\rule{0pt}{.8ex+\baselineskip}}{$Z(\to c\bar{c}) j \nu_e$} &  \\ \cline{1-5}
			\parbox{0pt}{\rule{0pt}{.8ex+\baselineskip}}{1.3} & 
			\parbox{0pt}{\rule{0pt}{.8ex+\baselineskip}}{0.145} & 
			\parbox{0pt}{\rule{0pt}{.8ex+\baselineskip}}{0.145} & 
			\parbox{0pt}{\rule{0pt}{.8ex+\baselineskip}}{0.111} & 
			\parbox{0pt}{\rule{0pt}{.8ex+\baselineskip}}{$6.9\times10^{-4}$} & \\
			{3.46} & 0.090 & 0.090 & 0.065 & $3.1\times10^{-4}$ & \\ 
			\cline{1-5}
			\parbox{0pt}{\rule{0pt}{.8ex+\baselineskip}}{$\sqrt s$ [TeV]} & 
			\parbox{0pt}{\rule{0pt}{.8ex+\baselineskip}}{${\gamma/g}^{*}(\to b\bar{b}) j \nu_e$} & 
			\parbox{0pt}{\rule{0pt}{.8ex+\baselineskip}}{${\gamma/g}^{*}(\to c\bar{c}) j \nu_e$} & 
			\parbox{0pt}{\rule{0pt}{.8ex+\baselineskip}}{${Z/\gamma/g}^{*}(\to j_{\ell}\bar{j_{\ell}}) j \nu_e$} & 
			\parbox{0pt}{\rule{0pt}{.8ex+\baselineskip}}{$\bar{t} j \nu_e$} \\
			\cline{1-5}  
			\parbox{0pt}{\rule{0pt}{.8ex+\baselineskip}}{1.3} & 
			\parbox{0pt}{\rule{0pt}{.8ex+\baselineskip}}{0.015} & 
			\parbox{0pt}{\rule{0pt}{.8ex+\baselineskip}}{$4.9\times10^{-4}$} & 
			\parbox{0pt}{\rule{0pt}{.8ex+\baselineskip}}{$4.8\times10^{-5}$} & 
			\parbox{0pt}{\rule{0pt}{.8ex+\baselineskip}}{0.052}  \\
			3.46 & $9.2\times10^{-3}$ & $2.3\times10^{-4}$ & $4.7\times10^{-5}$ & 0.025 \\ 
		\end{tabular}
	\end{center}  
	\caption{\small Event selection efficiencies corresponding to the center-of-mass energies of 1.3 and 3.46 TeV obtained by applying all the selection cuts to the signal and background events. The presented signal efficiencies correspond to the benchmark point $T_R=0.1$, $T_I=0$, as an example.}
	\label{table:eff} 
\end{table} 
\begin{itemize}
\item{$M_{b_{1}b_{2}}$: invariant mass of the two $b$-tagged jets,}
\item $\alpha_{b_1 b_2}^{\rm COM}$: angle between the momentum vectors of the $b$-tagged jets as measured in the center-of-momentum (COM) frame of the $b$-jets, 
\item{$\eta_{jet}$: pseudorapidity of the non-$b$-tagged jet,}
\item{$\Delta R_{\, b_1 b_2}$: distance between the two $b$-tagged jets,}
\item{$\Delta R_{\, b_1 j}$: distance between the leading $b$-tagged jet and the non-$b$-tagged jet,}
\item{$\Delta R_{\, b_2 j}$: distance between the sub-leading $b$-tagged jet and the non-$b$-tagged jet,}
\item{$\cos \alpha_{b_1 b_2}$: cosine of the angle between the momentum vectors of two $b$-tagged jets,}
\item $\slashed{E}_T$: missing transverse energy,
\item $H_T$: scalar sum of transverse momenta of all objects reconstructed in the detector.
\end{itemize}
\begin{figure*}[!h]
  \centering  
    \begin{subfigure}[b]{0.49\textwidth} 
    \centering
    \includegraphics[width=\textwidth]{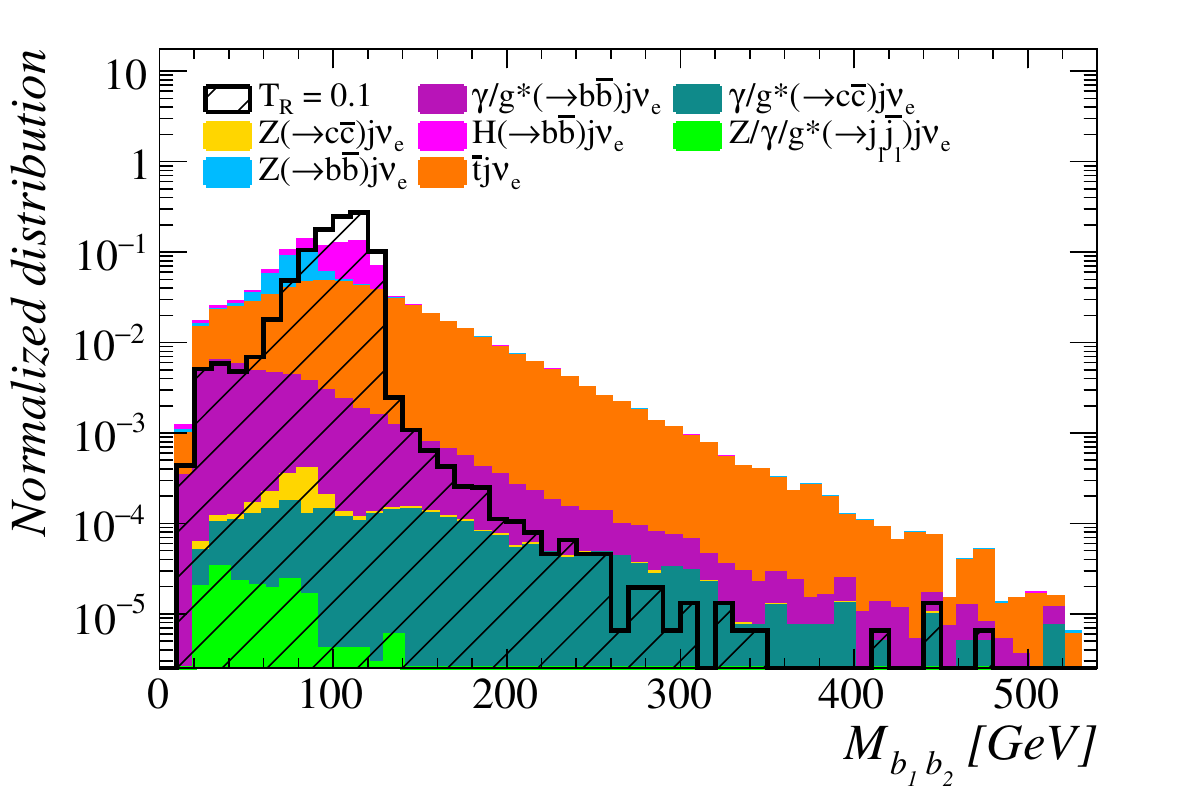}
    \caption{}
    \label{dist-bjetsInvariantMass}
    \end{subfigure} 
    \begin{subfigure}[b]{0.49\textwidth}
    \centering
    \includegraphics[width=\textwidth]{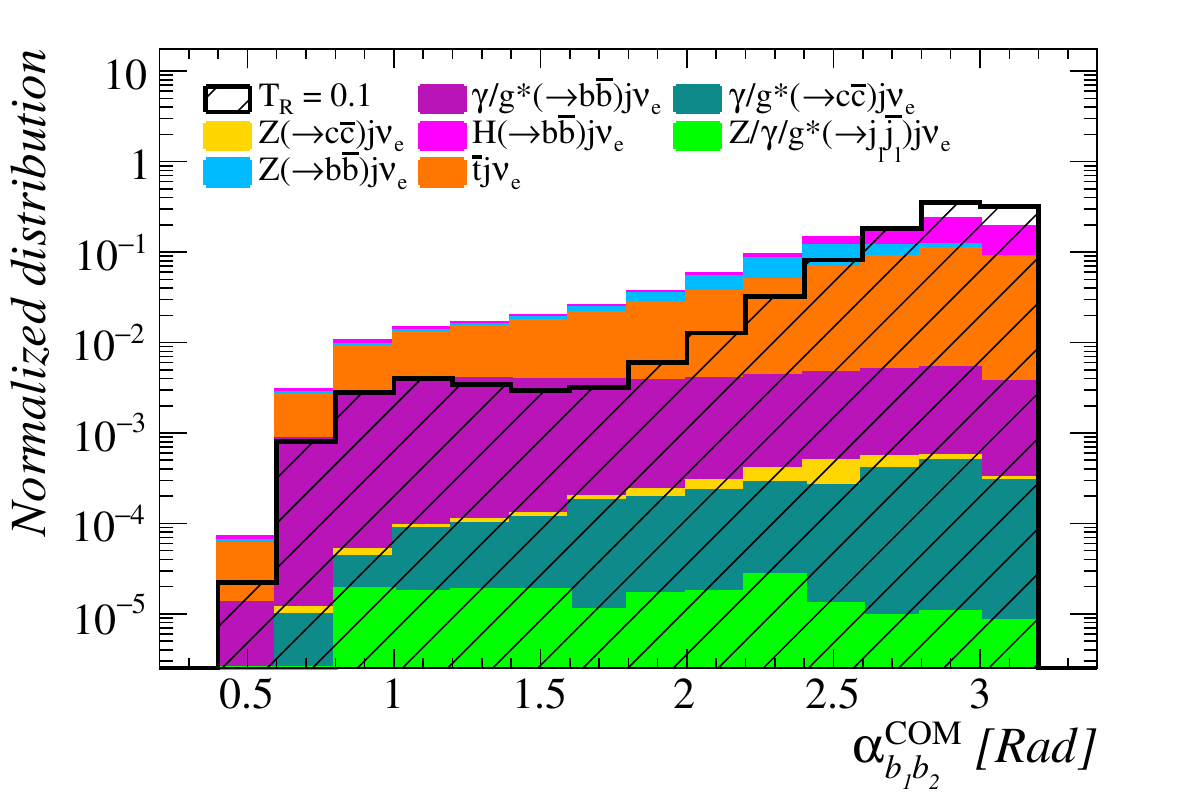}
    \caption{}
    \label{dist-alphaZMF}
    \end{subfigure} 
    \begin{subfigure}[b]{0.49\textwidth} 
    \centering
    \includegraphics[width=\textwidth]{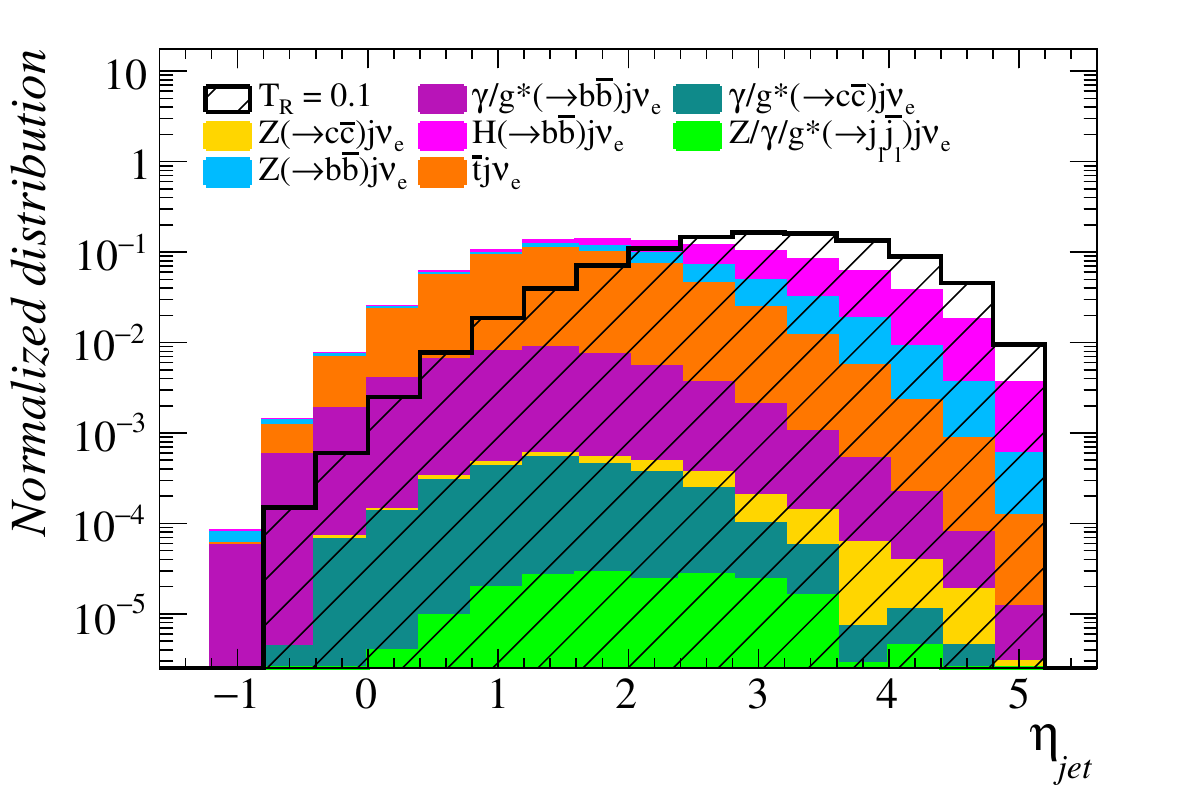}
    \caption{}
    \label{dist-jet_eta}
    \end{subfigure}
    \begin{subfigure}[b]{0.49\textwidth} 
    \centering
    \includegraphics[width=\textwidth]{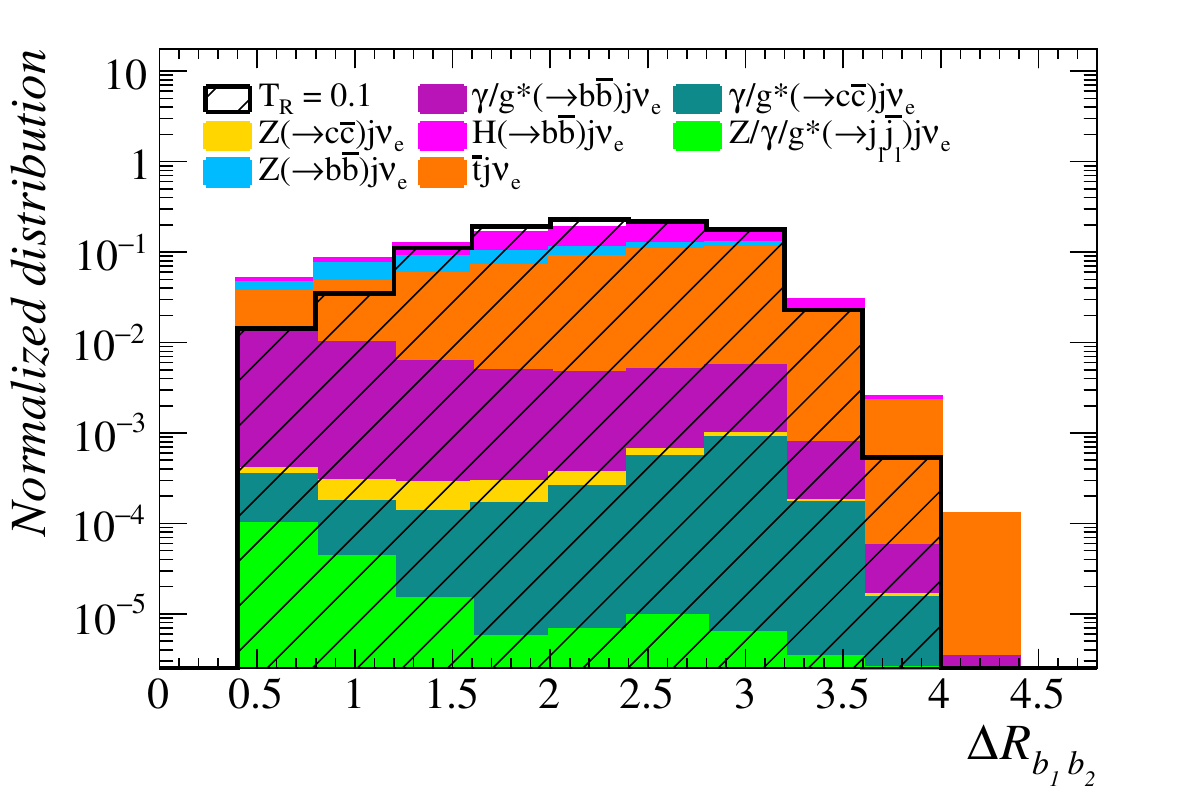}
    \caption{}
    \label{dist-deltaR_bjets}
    \end{subfigure} 
    \begin{subfigure}[b]{0.49\textwidth}
    \centering
    \includegraphics[width=\textwidth]{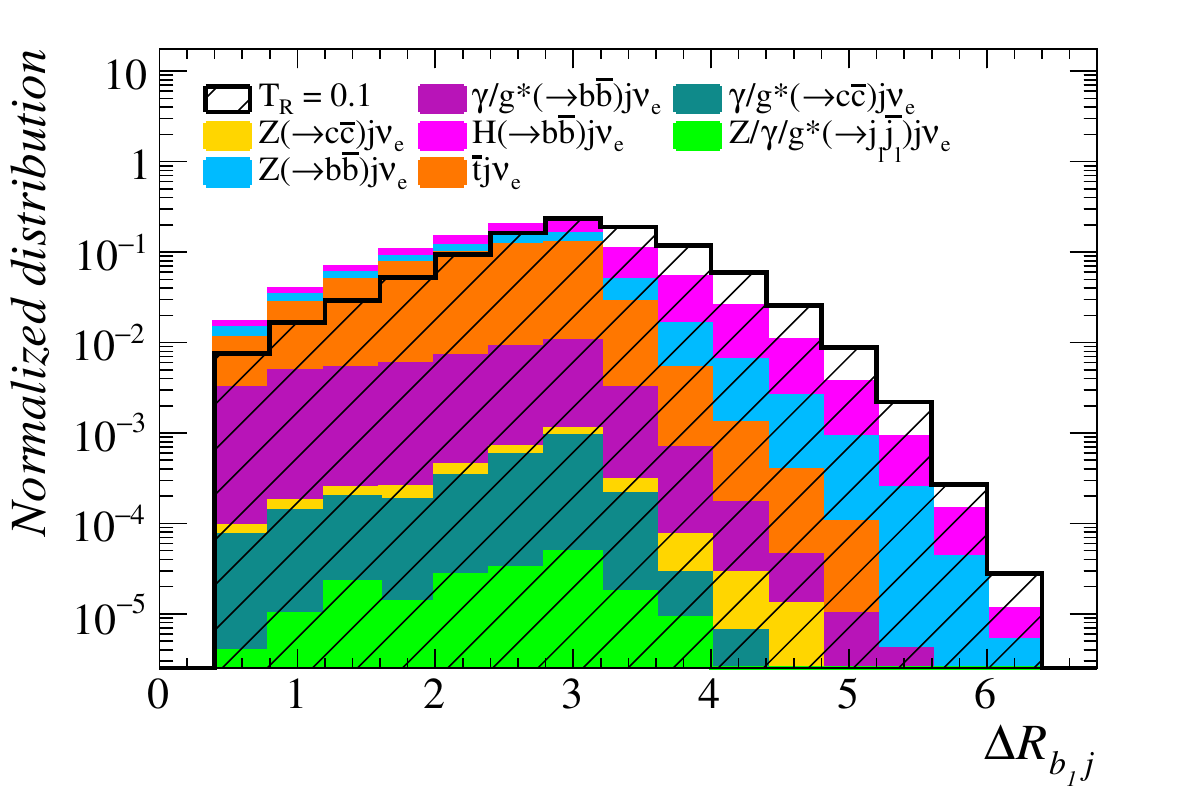}
    \caption{}
    \label{dist-deltaR_bjets}
    \end{subfigure}
    \begin{subfigure}[b]{0.49\textwidth}
    \centering
    \includegraphics[width=\textwidth]{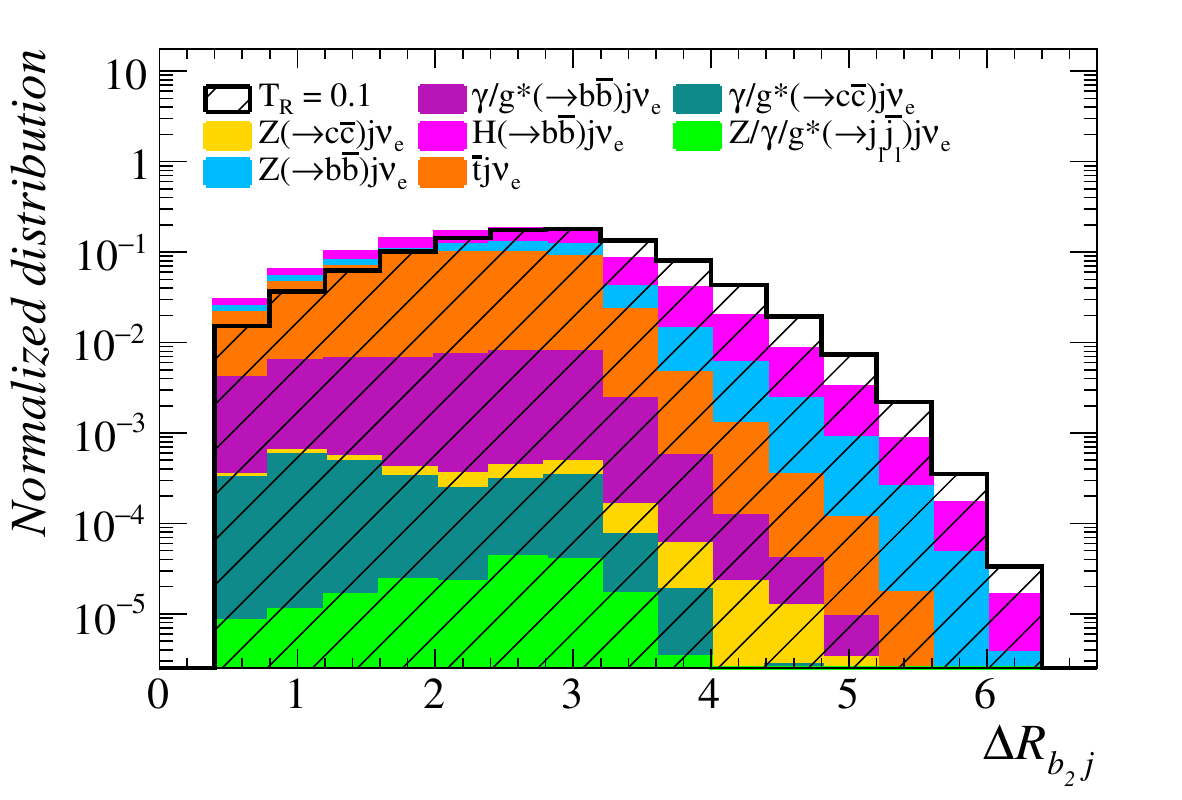}
    \caption{}
    \label{dist-deltaR_bjets}
    \end{subfigure}
\caption{\small Distributions of the discriminating variables obtained for the signal and background processes at the center-of-mass energy of 1.3 TeV corresponding to the benchmark point $T_R=0.1$, $T_I=0$. The distributions are normalized to unity.}
\label{fig:MVAinput-1}
\end{figure*}
\begin{figure*}[!ht]\ContinuedFloat
    \centering
    \begin{subfigure}[b]{0.49\textwidth}
    \centering
    \includegraphics[width=\textwidth]{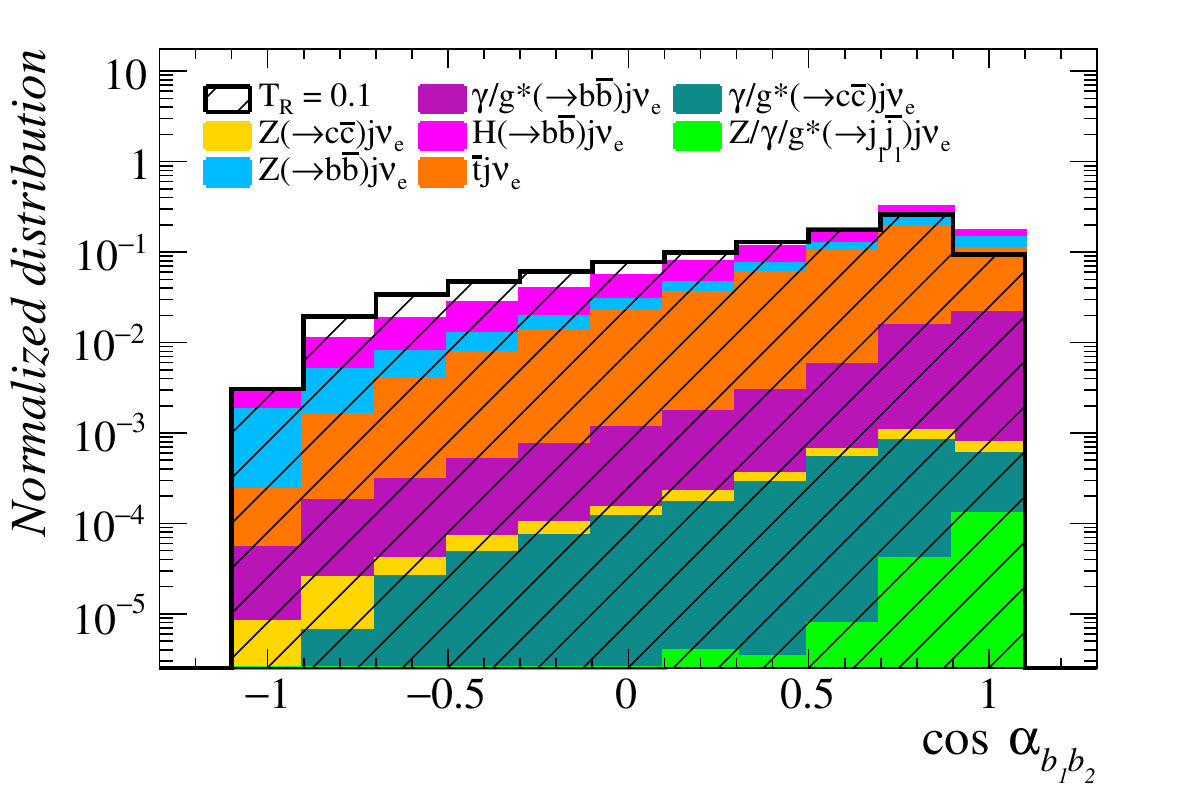}
    \caption{}
    \label{dist-deltaR_bjets}
    \end{subfigure} 
    \begin{subfigure}[b]{0.49\textwidth}
    \centering
    \includegraphics[width=\textwidth]{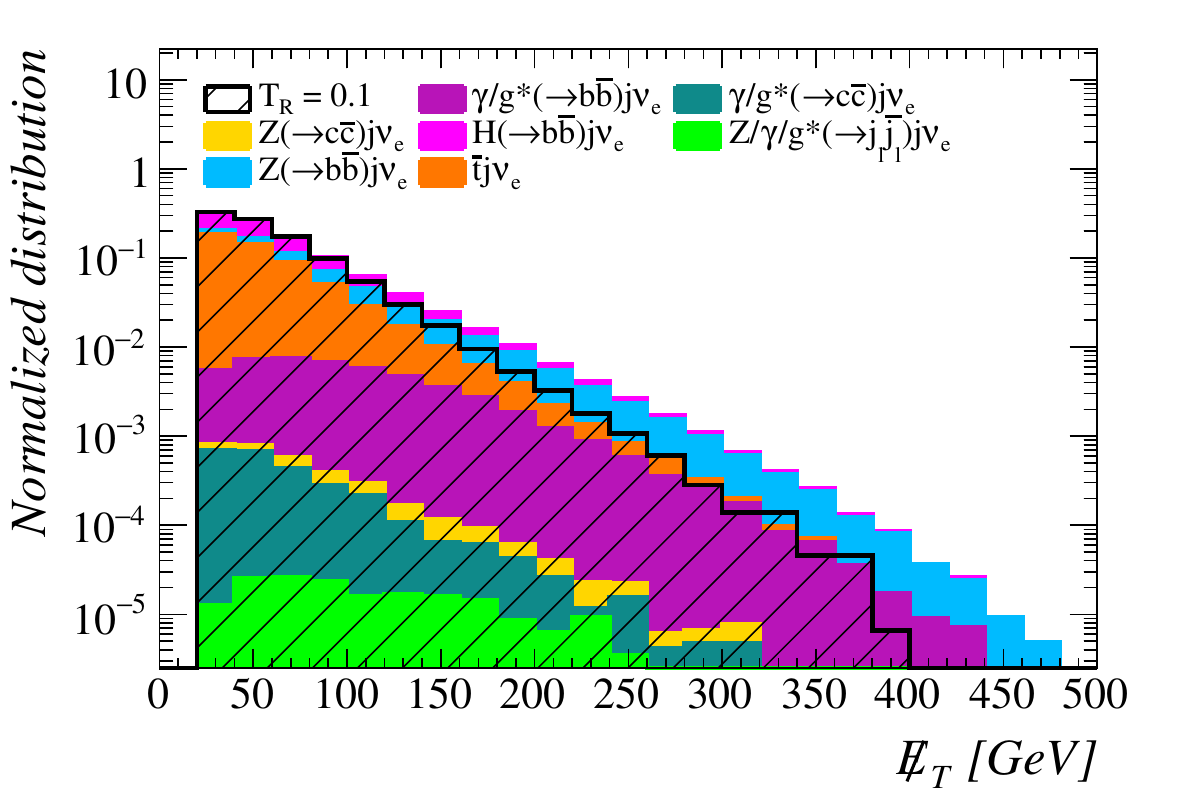}
    \caption{}
    \label{dist-deltaR_bjets}
    \end{subfigure}
    \begin{subfigure}[b]{.49\textwidth}
    \includegraphics[width=\textwidth]{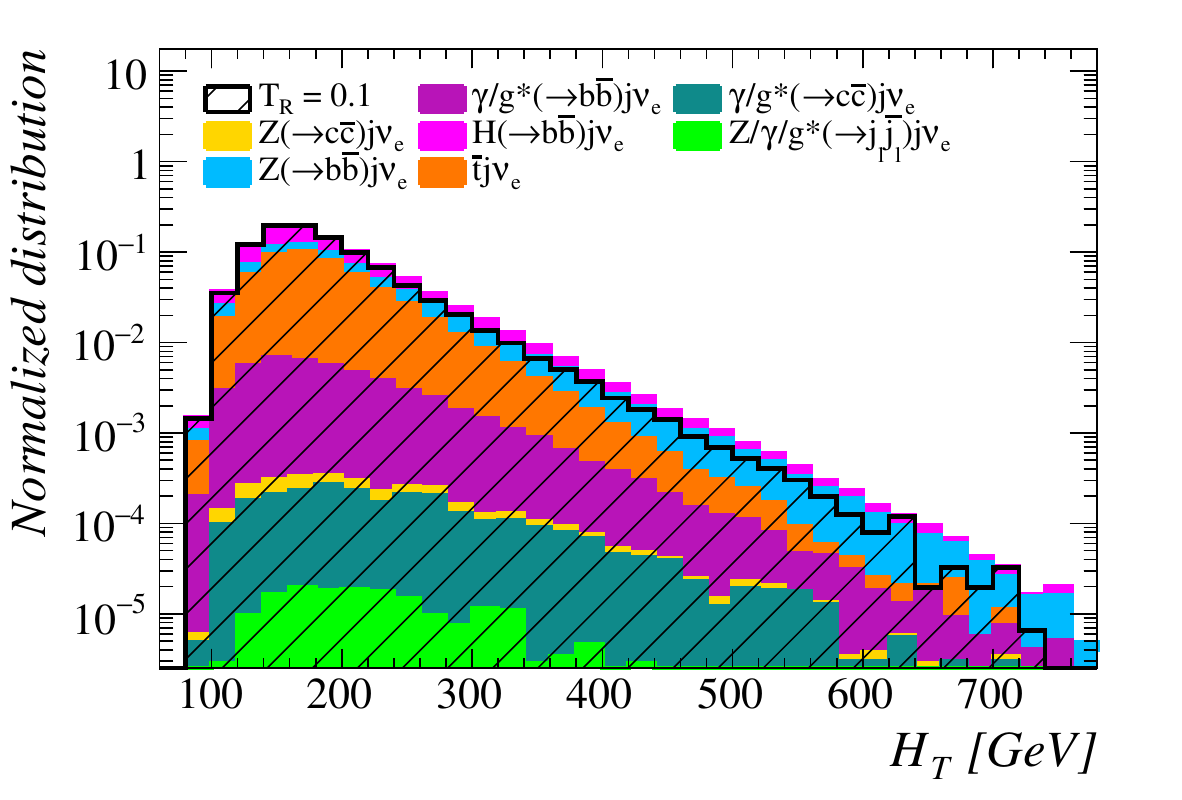}
    \caption{}
    \label{dist-deltaR_bjets}
    \end{subfigure}
\caption{\small Distributions of the discriminating variables obtained for the signal and background processes at the center-of-mass energy of 1.3 TeV corresponding to the benchmark point $T_R=0.1$, $T_I=0$. The distributions are normalized to unity.}
\label{fig:MVAinput-1}
\end{figure*}
Fig. \ref{fig:MVAinput-1} shows the distributions of the discriminating variables obtained for the signal and background processes assuming the center-of-mass energy of 1.3 TeV and the benchmark point $T_R=0.1$, $T_I=0$, as an example. The $b\bar{b}$ invariant mass variable, $M_{b_{1}b_{2}}$, is mostly effective in suppressing the background processes in which the $b\bar{b}$ pair does not originate from the Higgs boson. The $b\bar{b} j \nu_e$ production, where $b\bar{b}$ originates from a photon, a gluon or a $Z$ boson, is an example of such processes (see Fig. \ref{dist-bjetsInvariantMass}). The decay of the Higgs boson into a pair of $b$ quarks in the signal process is a two-body decay. The decay products are, therefore, produced in a back-to-back configuration in the rest frame of the decaying Higgs boson. As a result, the distribution of the angle $\alpha_{b_1 b_2}^{\rm COM}$ peaks at $\pi$ Radians for the signal process. This is, however, not the case for the background processes in which the $b\bar{b}$ pair does not originate from the Higgs boson. This variable is, therefore, advantageous and can reduce such background processes as can be seen from Fig. \ref{dist-alphaZMF}. The pseudorapidity of the non-$b$-tagged jet, $\eta_{jet}$, is also a useful discriminant since the non-$b$-tagged jet in the signal process is likely to be reconstructed in the forward region because of the nature of $t$-channel processes. For most of the background processes, the behavior of this variable is different from the signal as seen in Fig. \ref{dist-jet_eta}. The signal process is the same as the background process $H(\to b\bar{b}) j \nu_e$ except that it originates from dimension-six $Hb\bar{b}$ couplings. As a result, the signal events have the same kinematics as the $H(\to b\bar{b}) j \nu_e$ background events leading to the same distributions in Fig. \ref{fig:MVAinput-1}.
 
As seen in the $b\bar{b}$ invariant mass distributions, Fig. \ref{dist-bjetsInvariantMass}, there is a discrepancy between the signal peak and the nominal Higgs boson mass ($\approx125$ GeV). This is also the case for the nominal $Z$ boson mass ($\approx 91$ GeV) and the peak of the $b\bar{b}$ invariant mass distribution obtained for the background process $Z(\to b\bar{b}) j \nu_e$. This discrepancy mainly arises from the missing neutrinos produced in the decays of unstable hadrons inside jets, uncertainties in the jet reconstruction algorithm, uncertainties in the measurement of the energy of the reconstructed objects, etc.

\begin{figure*}[t]
\begin{center}
\resizebox{0.53\textwidth}{!}{\includegraphics{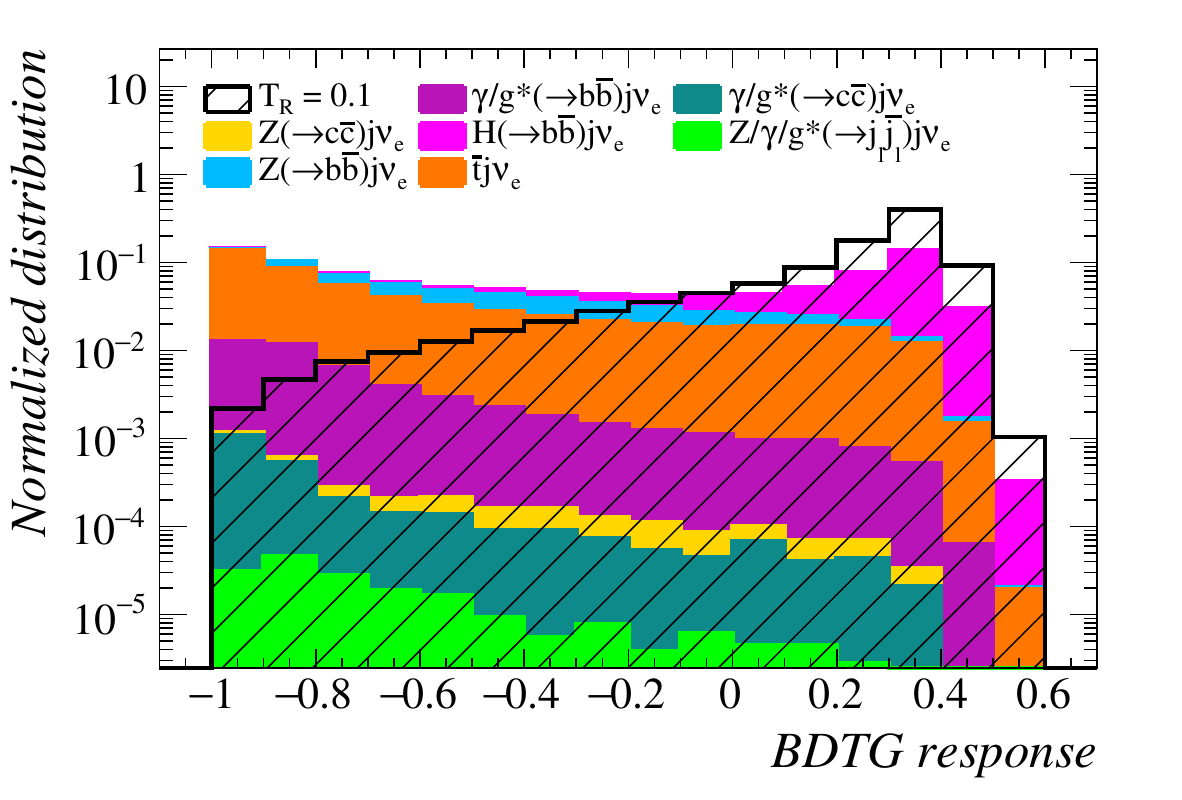}}
\caption{{\small 
BDTG response obtained for the signal and background processes corresponding to the center of mass energy of 1.3 TeV and the benchmark point $T_R=0.1$, $T_I=0$. The distributions of the signal and total background are separately normalized to unity.
} \label{fig:BDTG}}
\end{center}
\end{figure*}

To best discriminate the signal from background, a multivariate technique using the TMVA (Toolkit for Multivariate Data Analysis) package \cite{Hocker:2007ht,Stelzer:2008zz,Speckmayer:2010zz,Therhaag:2010zz} is deployed. All algorithms available in the TMVA package are examined using the receiver operating characteristic (ROC) curve to find the algorithm with the highest discrimination power. Based on the obtained ROC curves, the gradient Boosted Decision Trees (BDTG) algorithm \cite{Xia:2018cfz} gives the best signal-background discrimination and is, therefore, used in this analysis. The distributions of the discriminating variables obtained for the signal and background processes are passed to the BDTG algorithm as input. All background processes are considered in the training stage according to their respective weights. The same sets of variables are used for the two assumed collider energy scenarios. To ensure that overtraining does not occur, the TMVA overtraining check is performed. The BDTG response obtained for the test and training samples are compared by performing the Kolmogorov-Smirnov (K-S) test for both the signal and background processes. With a reasonable value of the K-S test, the consistency of the responses obtained for the training and test samples is ensured. Fig. \ref{fig:BDTG} shows the BDTG response distributions obtained for the signal and background processes assuming the benchmark point $T_R=0.1$, $T_I=0$, and the center-of-mass energy of 1.3 TeV. 
As seen, the signal is mostly overwhelmed by the SM $H(\to b\bar{b}) j \nu_e$ background process. This process is the most severe background process as it is the SM counterpart of the signal. The signal is mildly affected by the $\bar{t} j \nu_e$ and $Z(\to b\bar{b}) j \nu_e$ background processes as these processes are well-separated from the signal. Other background processes including ${\gamma/g}^{*}(\to b\bar{b}) j \nu_e$, ${\gamma/g}^{*}(\to c\bar{c}) j \nu_e$, $Z(\to c\bar{c}) j \nu_e$ and ${Z/\gamma/g}^{*}(\to j_{\ell}\bar{j_{\ell}}) j \nu_e$ are less important and well under control because of the efficient signal-background discrimination and the small contribution to the total background. According to the TMVA output, the variables $M_{b_1 b_2}$, $\alpha_{b_1 b_2}^{\rm COM}$ and $\eta_{jet}$ are the most powerful variables to discriminate the signal from the background in this analysis (see Figs. \ref{dist-bjetsInvariantMass},\ref{dist-alphaZMF},\ref{dist-jet_eta}). The obtained BDTG response is used to constrain the dimension-six $Hb\bar{b}$ couplings at the LHeC and FCC-he.

\section{Results and discussions}
\label{sec:Results}

To obtain the expected limits on $T_I$ and $T_R$, the CL$_{s}$ method \cite{cl1,cl2} is used. In this method, the log-likelihood functions $\mathcal{L}_{bkg}$ and $\mathcal{L}_{signal+bkg}$ for the background and signal+background hypotheses are constructed as the multiplication of Poissonian likelihood functions. The $p$-values for the background and signal+background hypotheses are computed using the log-likelihood functions ratio, $Q = 2 \ln(\mathcal{L}_{signal+bkg}/\mathcal{L}_{bkg})$. To find the limits at $95\%$ CL, the signal cross section is constrained using the condition:
\begin{eqnarray}
 \text{CL}_{s} = \frac{\mathcal{P}_{signal+bkg} (Q > Q_{0})}{1- \mathcal{P}_{bkg} (Q < Q_{0})} \leq 0.05,
 \end{eqnarray}
where $Q_0$ denotes the expected value of the test statistics $Q$. The CL$_s$ calculations are performed using the {\tt RooStats} package \cite{Moneta:2010pm}. 
The limits obtained in this analysis can be affected by systematic uncertainties arising from the jet reconstruction algorithm, $b$-tagging technique, measurement of the energy and momentum of particles, etc. To take into account potential systematic uncertainties, an overall uncertainty of $10\%$ on the event selection efficiencies obtained for the signal and background processes is considered. The limits are computed assuming the integrated luminosities of 1, 2, and 10 ab$^{-1}$. 

\begin{figure*}[!t]
\begin{center}
\resizebox{0.7\textwidth}{!}{\includegraphics{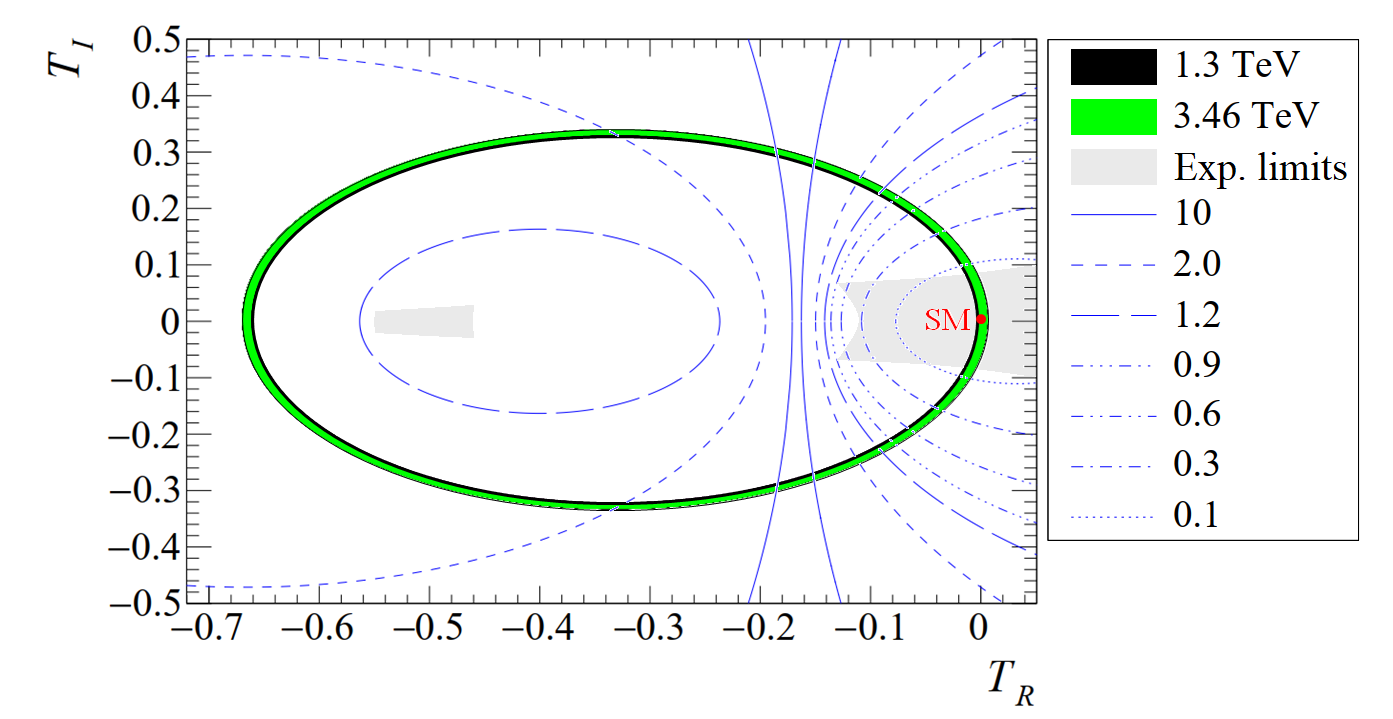}}
\caption{\small Expected $95\%$ CL exclusion limits in the $T_I \mhyphen T_R$ plane obtained for the energy scenarios 1.3 TeV (LHeC) and 3.46 TeV (FCC-he) and the integrated luminosity of 1 ab$^{-1}$. The provided limits are obtained assuming an overall uncertainty of $10\%$ on the signal and background event selection efficiencies. The red point shows the SM. {Also shown are the contour lines showing the value of the ratio $\mathcal{R}$ at a given point of the parameter space. The gray regions show the current experimental $95\%$ CL exclusion limits taken from Ref. \cite{Fuchs:2020uoc}.}} 
\label{limits_13TeV_34TeV}
\end{center}
\end{figure*}

Fig. \ref{limits_13TeV_34TeV} shows the expected 95$\%$ CL exclusion limits in the $T_I \mhyphen T_R$ plane corresponding to the energy scenarios 1.3 and 3.46 TeV and the integrated luminosity of 1 $ab^{-1}$.
The provided limits have been computed assuming that the kinematics of the signal events does not depend on the values of the $T_I$ and $T_R$ coupling constants. A comparison shows that the limits are slightly improved when the center-of-mass energy increases from 1.3 TeV to 3.46 TeV. 
\begin{figure*}[!ht]
  \centering  
    \begin{subfigure}[b]{0.49\textwidth}
    \centering
    \includegraphics[width=\textwidth]{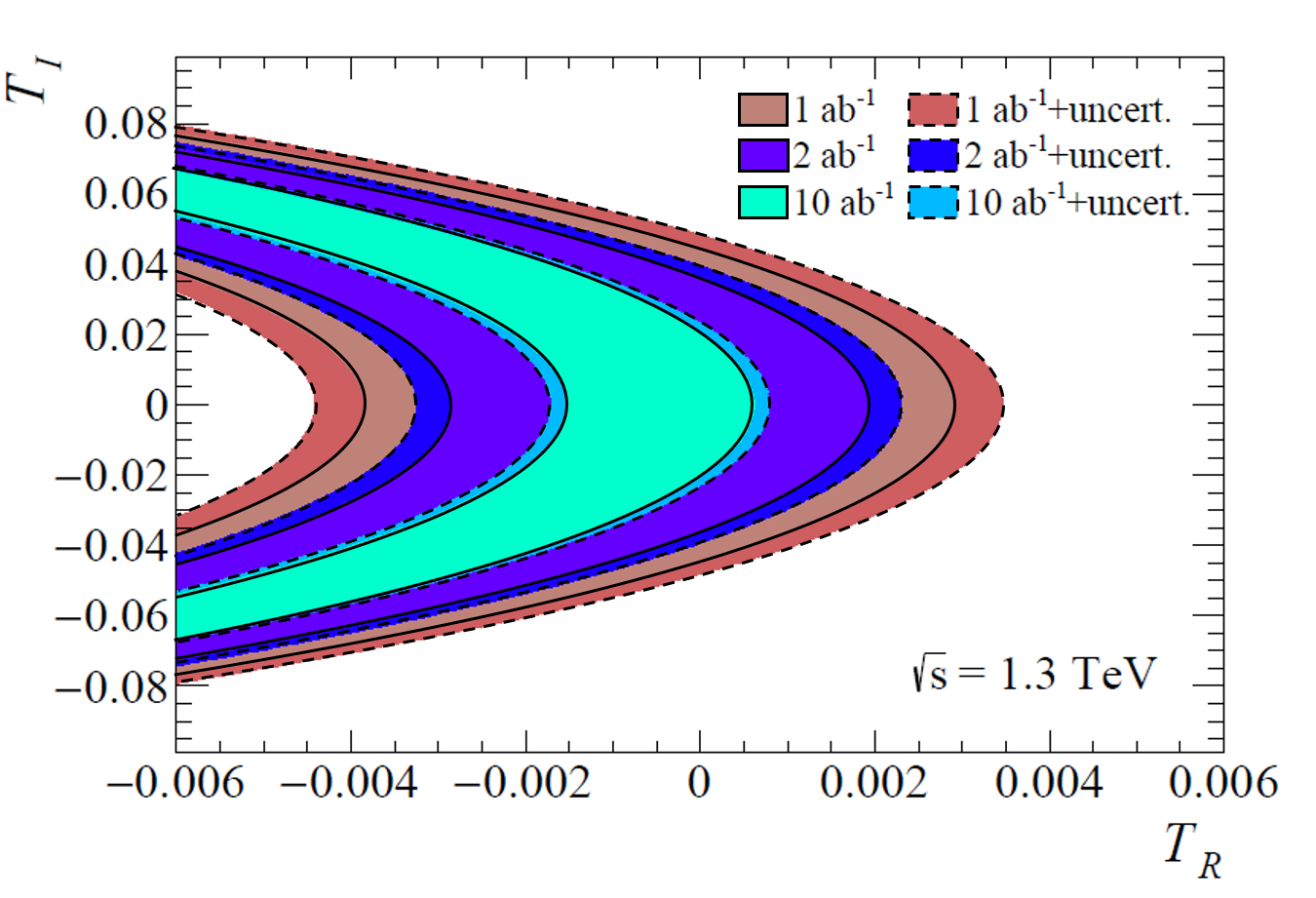}
    \caption{}
    \label{limits_right13}
    \end{subfigure} 
    \begin{subfigure}[b]{0.49\textwidth} 
    \centering
    \includegraphics[width=\textwidth]{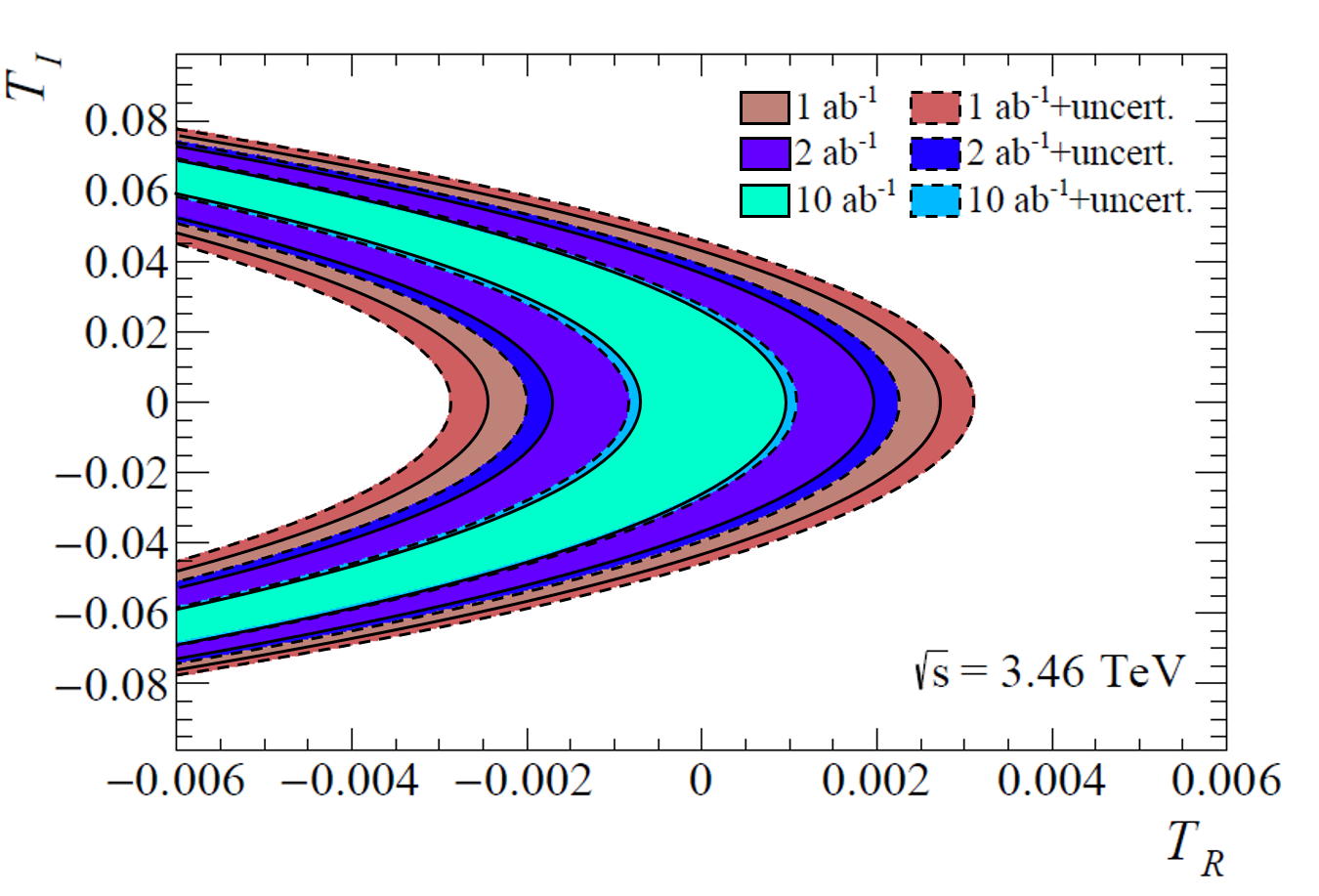}
    \caption{}
    \label{limits_right34}
    \end{subfigure}
    \begin{subfigure}[b]{0.49\textwidth} 
    \centering
    \includegraphics[width=\textwidth]{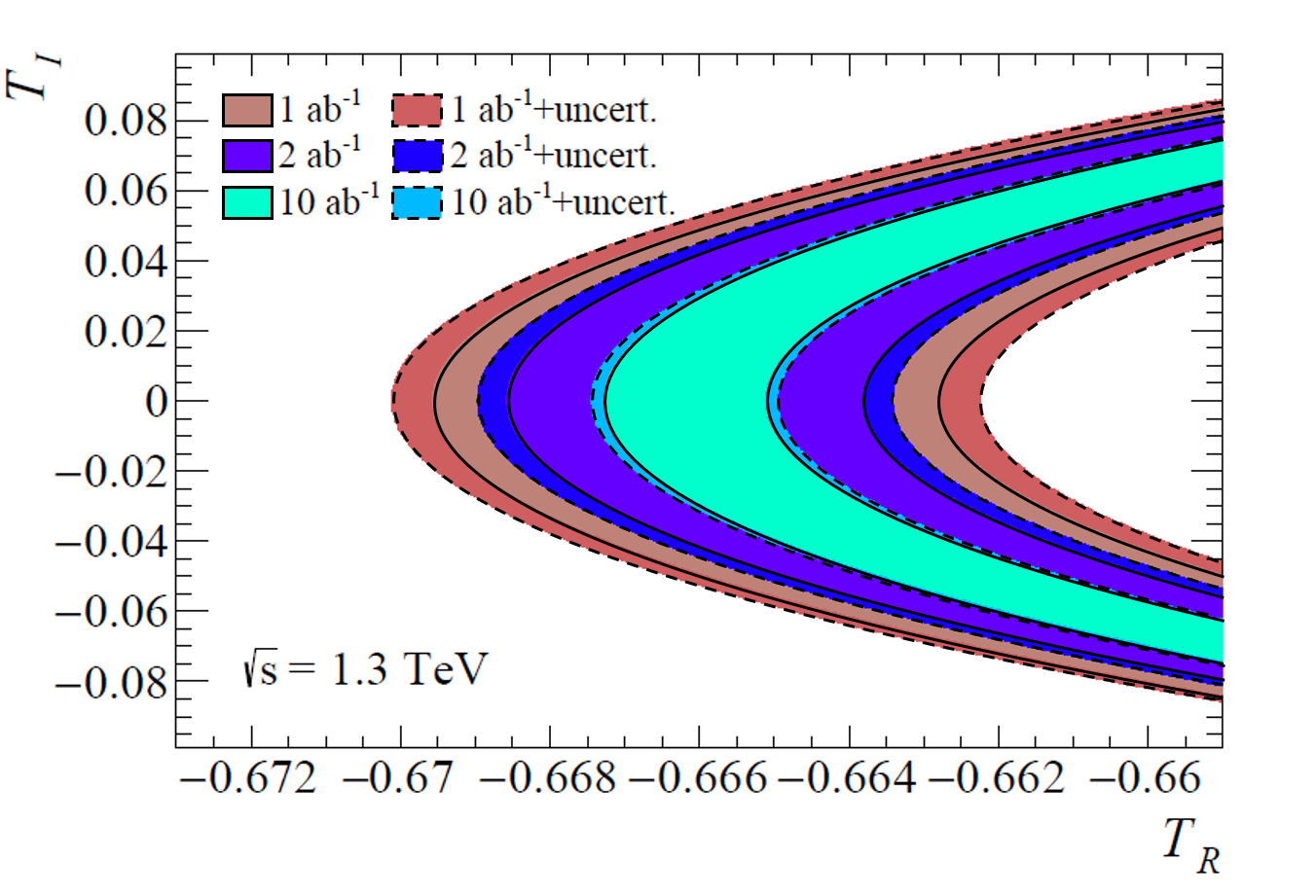}
    \caption{}
    \label{limits_left13}
    \end{subfigure} 
    \begin{subfigure}[b]{0.49\textwidth}
    \centering
    \includegraphics[width=\textwidth]{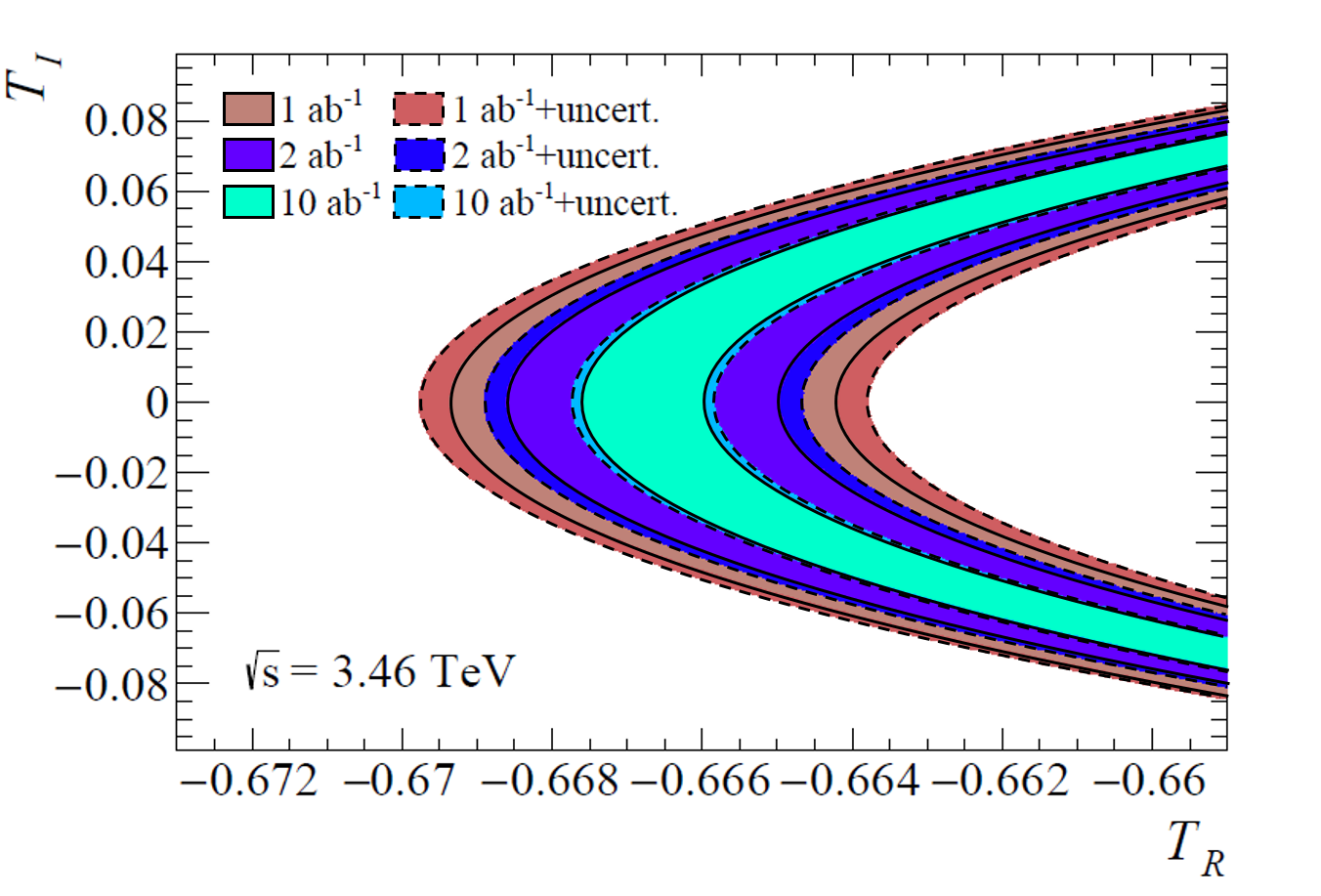}
    \caption{}
    \label{limits_left34}
    \end{subfigure}
\caption{\small Expected $95\%$ CL exclusion limits in two sub-regions of the $T_I \mhyphen T_R$ parameter space obtained for the energy scenarios 1.3 TeV (LHeC) and 3.46 TeV (FCC-he) and the integrated luminosities 1, 2 and 10 ab$^{-1}$. Exclusion limits obtained assuming an overall uncertainty of $10\%$ on the signal and background event selection efficiencies are also shown (dashed lines).}
\label{fig:limits}
\end{figure*}
{The two gray regions in this figure show the current experimental $95\%$ CL exclusion limits. It is seen that the limits obtained in the present study significantly improve the current experimental limits. Most of the larger (gray) region on the right side and the whole of the smaller region on the left side are excluded by the obtained limits. Also shown in this figure are the contour lines showing the value of the ratio $\mathcal{R}$ (contours are the same as the contours in Fig. \ref{xsec2dim_plot}). The dotted, dash-dotted and dash-double-dotted lines respectively correspond to the $\mathcal{R}$ values 0.1, 0.3 and 0.6, and at any point outside the dash-double-dotted contour line, $\mathcal{R}$ is found to be larger than 0.6. As discussed in section \ref{sec:Theoretical}, the contribution of the $1/\Lambda^4$ corrections to the signal cross section is small in the parameter space regions around the zero point ($T_R=T_I=0$). However, in regions far from the zero point, the $1/\Lambda^4$ corrections are large and non-negligible. Consequently, the EFT approach deployed in this study and the obtained limits are only valid in regions around the zero point. In other regions, the dimension-eight operators may also induce non-negligible $1/\Lambda^4$ corrections and thus should be taken into account. Assuming that $1/\Lambda^4$ corrections induced by dimension-eight operators (if considered) are similar in size to the $1/\Lambda^4$ corrections induced by the dimension-six operators, the parameter space region enclosed by the dotted contour ($\mathcal{R}\lesssim 0.1$) can be roughly considered to be compatible with the systematic uncertainty assumed in this section to compute the limits, and therefore, the obtained exclusion limits in this region are valid. It is worth mentioning that, as seen in Fig. \ref{limits_13TeV_34TeV}, most of the region with $\mathcal{R}>0.1$ has already been excluded by the current experimental limits and is of no interest to us. In fact, a large portion of the unexcluded (gray) region, which is the main focus of this study, is covered by the (valid) $\mathcal{R}\lesssim 0.1$ region.}

For a better comparison of the limits obtained at different center-of-mass energies (and integrated luminosities), Fig. \ref{fig:limits} shows the exclusion limits in two sub-regions of the parameter space. The limits in these figures correspond to the integrated luminosities of 1, 2 and 10 $ab^{-1}$. As seen, the $T_R$ limits are asymmetric. This can be understood as an effect of the interference involving the CP-even dimension-six operators affecting the signal process. Furthermore, the signal cross section is highly sensitive to $T_R$ because of such interferences, which results in more stringent limits for $T_R$ compared with the limits obtained for $T_I$.
Comparing the expected limits obtained in this analysis with the current experimental limits on $T_R$ and $T_I$ shows a significant improvement in the limits in this analysis. To be more specific, the obtained expected limits and the current experimental limits on $T_R$ and $T_I$ assuming one non-zero coupling constant at a time are provided in Tab. \ref{table:result}.
\begin{table}[!t]
\begin{subtable}{1\textwidth}
\centering
\begin{tabular}{cccc}
 $\sqrt{s}$ [TeV] & $\mathcal{L}$ [ab$^{-1}$] & $T_R$ & $T_I$ \\
\cline{1-4}
\multirow{6}{*}{1.3} & 
\multirow{2}{*}{1} 
& $[-0.6697,-0.6629]\cup[-0.0037,0.0031]$ & $[-0.0460,0.0460]$\\
& & ($[-0.6702,-0.6624]\cup[-0.0043,0.0036]$) & ($[-0.0499,0.0499]$)\\
\cline{2-4}
& \multirow{2}{*}{2} 
& $[-0.6687,-0.6639]\cup[-0.0027,0.0021]$ & $[-0.0381,0.0381]$\\
& & ($[-0.6691,-0.6635]\cup[-0.0031,0.0025]$) & ($[-0.0414,0.0414]$)\\
\cline{2-4}
& \multirow{2}{*}{10} 
& $[-0.6674,-0.6653]\cup[-0.0014,0.0008]$ & $[-0.0238,0.0238]$\\
& & ($[-0.6676,-0.6651]\cup[-0.0016,0.0009]$) & ($[-0.0261,0.0261]$) \\
\cline{1-4}
\multirow{6}{*}{3.46} & 
\multirow{2}{*}{1} 
& $[-0.6690,-0.6639]\cup[-0.0026,0.0025]$ & $[-0.0409,0.0409]$\\
& & ($[-0.6695,-0.6635]\cup[-0.0030,0.0029]$) & ($[-0.0442,0.0442]$)\\
\cline{2-4}
& \multirow{2}{*}{2} 
& $[-0.6683,-0.6647]\cup[-0.0019,0.0017]$ & $[-0.0342,0.0342]$\\
& & ($[-0.6686,-0.6644]\cup[-0.0022,0.0020]$) & ($[-0.0370,0.0370]$) \\
\cline{2-4}
& \multirow{2}{*}{10} 
& $[-0.6673,-0.6657]\cup[-0.0009,0.0007]$ & $[-0.0224,0.0224]$\\
& & ($[-0.6674,-0.6655]\cup[-0.0010,0.0009]$) & ($[-0.0243,0.0243]$)\\
\cline{1-4} 
\end{tabular}
\caption{}
\label{table:result1}
\end{subtable}
\begin{subtable}{1\textwidth}
\centering
\medskip
\begin{tabular}{ccc}
 &  $T_R$ & $T_I$ \\
\cline{1-3}
$\mu_{b\bar{b}}$ & $[-0.55,-0.46]\cup[-0.12,0.33]$ & $[-0.43,0.43]$  \\
$d_{e}$ & $-$ & $[-0.09,0.09]$ \\
\end{tabular}
\caption{}
\label{table:result2}
\end{subtable}
\caption{Expected $95\%$ CL limits on $T_I$ and $T_R$ assuming one non-zero coupling constant at a time: a) Expected limits corresponding to the energy scenarios 1.3 and 3.46 TeV and the integrated luminosities 1, 2 and 10 ab$^{-1}$. Limits in the parentheses are obtained assuming an overall uncertainty of $10\%$ on the signal and background event selection efficiencies, b) Current experimental limits derived from the Higgs boson signal strength $\mu_{b\bar{b}}$ and the electron EDM data taken from Ref. \cite{Fuchs:2020uoc}.} 
\label{table:result}
\end{table} 
As seen in Tab. \ref{table:result2}, $T_R$ is currently constrained by $T_{R} \in [-0.55,-0.46]\cup[-0.12,0.33]$ derived from the Higgs boson signal strength $\mu_{b\bar{b}}$ data \cite{Fuchs:2020uoc}. A comparison with the expected limits provided in Tab. \ref{table:result1} shows that the first sub-range, $[-0.55,-0.46]$, is totally excluded, and the second sub-range, $[-0.12,0.33]$, is tightened by about two (three) orders of magnitude at the integrated luminosity of 1 (10) $ab^{-1}$ in this analysis. As seen in Tab. \ref{table:result1}, similar expected limits are obtained at the two examined energy scenarios with slightly stronger limits for the higher energy, $\sqrt{s}=3.46$ TeV. As can be seen, the most stringent limit on $T_{R}$ obtained in this study is $T_{R} \in [-0.6674,-0.6655]\cup[-0.0010,0.0009]$ which corresponds to the center-of-mass energy of 3.46 TeV and the integrated luminosity of 10 $ab^{-1}$. A comparison also shows an improvement in the limits on $T_{I}$ in this study. This improvement is not as large as the improvement made to the $T_{R}$ limits. It is, however, significant. The strongest experimental limit on $T_I$ ($T_{I} \in [-0.09 , 0.09]$) is currently derived from the electron EDM data \cite{Fuchs:2020uoc}. The limits on $T_I$ obtained in this analysis are (less than one order of magnitude) stronger than this limit. As seen, the strongest limit on $T_{I}$ obtained in this study is $T_{I} \in [-0.0243,0.0243]$ which corresponds to the center-of-mass energy of 3.46 TeV and the integrated luminosity of 10 $ab^{-1}$. It is seen that the FCC-he and LHeC are both capable of improving the current limits on $T_{R}$ and $T_{I}$ significantly with the help of the CC channel Higgs boson production. Slightly stronger limits at the higher center-of-mass energy, $\sqrt{s}=3.46$ TeV, makes FCC-he a slightly more powerful machine than LHeC (assuming the same integrated luminosities) to probe the $Hb\bar{b}$ dimension-six couplings.

\section{Summary and conclusions} 
\label{sec:Conclusion}
In this work, the capability of the future electron-proton colliders, FCC-he and LHeC, to probe the $Hb\bar{b}$ CP-even and CP-odd dimension-six couplings was studied. The NP coupling constants $T_R$ and $T_I$ were constrained by performing a search for the Higgs boson production through the CC channel with subsequent decay of the Higgs boson into a pair of bottom quarks, $e^-p \to Hj\nu_e, H \to b\bar{b}$. The signal and relevant background processes were simulated using a Monte Carlo event generator and detector effects were applied. To discriminate the signal from the background, a set of well-chosen discriminating variables were analyzed using a multivariate technique. It was seen that, among the available multivariate analysis algorithms, the BDTG algorithm gives the best signal-background discrimination. Using the BDTG output and with the help of the CL$_s$ technique, expected $95\%$ CL limits on the coupling constants $T_R$ and $T_I$ were computed and the exclusion limits in the $T_I \mhyphen T_R$ plane were provided. The provided limits correspond to the integrated luminosities of 1, 2 and 10 ab$^{-1}$ and the center-of-mass energies of 1.3 TeV (LHeC) and 3.46 TeV (FCC-he). A comparison between the obtained limits and the current experimental limits shows that a significant region of the unprobed $T_I \mhyphen T_R$ parameter space becomes accessible with the help of the present analysis. Assuming one non-zero coupling constant at a time to make a comparison, the $T_R$ parameter is currently limited to the range $T_{R} \in [-0.55,-0.46]\cup[-0.12,0.33]$ based on the Higgs signal strength data. It is seen by a comparison that the present study excludes the sub-range $[-0.55,-0.46]$ completely and tightens the sub-range $[-0.12,0.33]$ by at least two orders of magnitude. The improvement in the limits can also be of three orders of magnitude if an integrated luminosity of 10 $ab^{-1}$ is assumed. A comparison also shows that an improvement of less than one order of magnitude is possible for the current experimental limit on $T_I$, $T_{I} \in [-0.09 , 0.09]$, which is derived from the electron EDM data. The most stringent expected limits on $T_R$ and $T_I$ obtained in this study are $T_{R} \in [-0.6674,-0.6655]\cup[-0.0010,0.0009]$ and $T_{I} \in [-0.0243,0.0243]$ which correspond to the center-of-mass energy of 3.46 TeV and the integrated luminosity of 10 $ab^{-1}$. Based on the obtained results, the two center-of-mass energies assumed in this study result in similar limits with slightly stronger limits for the higher energy, $\sqrt{s}=3.46$ TeV. It is seen that both of the FCC-he and LHeC provide the possibility to improve the current limits on $T_R$ and $T_I$ significantly. Assuming the same integrated luminosities, the expected limits at FCC-he are slightly stronger due to its higher planned collision energy. Using 1 ab$^{-1}$ of data, the $T_R$ and $T_I$ parameters can be probed down to the orders of $10^{-3}$ and $10^{-2}$, respectively, at these future electron-proton colliders, providing the possibility for constraining the dimension-six operators affecting the $Hb\bar{b}$ coupling with unprecedented precision. {It should be noted that the validity of the bounds obtained in this study is limited to those regions of parameter space where the contribution of the $1/\Lambda^4$ corrections to the signal cross section is negligible. In regions with a large $1/\Lambda^4$ correction, an EFT description based solely on dimension-six operators is incomplete since dimension-eight operators may also modify the signal rate (kinematics) significantly through corresponding $1/\Lambda^4$ corrections.} It can be concluded that the present study can serve as a tool for searching for signals from higher dimensional operators relevant to the Higgs boson decay and provides a great opportunity to probe a significant region of the parameter space inaccessible to date. 

%
\section*{Acknowledgments}
The authors thank the School of Particles and Accelerators of IPM for financial support of this project. The authors thank S. Tizchang for the fruitful comments and helps. H. Khanpour is thankful to the Department of 
Physics at the University of Udine, and the University of Science and Technology of Mazandaran for the financial support provided for this project. M. Mohammadi Najafabadi is grateful to CERN TH-division for the nice hospitality and financial support. R. Jafari thanks the Iran Science Elites Federation for the financial support. 

%


\begin{thebibliography}{}

\bibitem{Chatrchyan:2012xdj} 
S.~Chatrchyan {\it et al.} [CMS Collaboration],
``Observation of a New Boson at a Mass of 125 GeV with the CMS Experiment at the LHC,''
Phys.\ Lett.\ B {\bf 716}, 30 (2012)
doi:10.1016/j.physletb.2012.08.021
[arXiv:1207.7235 [hep-ex]].

\bibitem{Aad:2012tfa} 
G.~Aad {\it et al.} [ATLAS Collaboration],
``Observation of a new particle in the search for the Standard Model Higgs boson with the ATLAS detector at the LHC,''
Phys.\ Lett.\ B {\bf 716}, 1 (2012)
doi:10.1016/j.physletb.2012.08.020
[arXiv:1207.7214 [hep-ex]].

\bibitem{r1}
J.~de Blas, M.~Cepeda, J.~D'Hondt, R.~K.~Ellis, C.~Grojean, B.~Heinemann, F.~Maltoni, A.~Nisati, E.~Petit and R.~Rattazzi, \textit{et al.}
JHEP \textbf{01}, 139 (2020)
doi:10.1007/JHEP01(2020)139
[arXiv:1905.03764 [hep-ph]].

\bibitem{CMS:2022dwd}
 [CMS],
Nature \textbf{607}, no.7917, 60-68 (2022)
doi:10.1038/s41586-022-04892-x
[arXiv:2207.00043 [hep-ex]].

\bibitem{ATLAS:2020fcp}
G.~Aad \textit{et al.} [ATLAS],
Eur. Phys. J. C \textbf{81}, no.2, 178 (2021)
doi:10.1140/epjc/s10052-020-08677-2
[arXiv:2007.02873 [hep-ex]].

\bibitem{hfull}
J.~de Blas, M.~Cepeda, J.~D'Hondt, R.~K.~Ellis, C.~Grojean, B.~Heinemann, F.~Maltoni, A.~Nisati, E.~Petit and R.~Rattazzi, \textit{et al.}
JHEP \textbf{01}, 139 (2020)
doi:10.1007/JHEP01(2020)139
[arXiv:1905.03764 [hep-ph]].


\bibitem{Fuchs:2020uoc}
E.~Fuchs, M.~Losada, Y.~Nir and Y.~Viernik,
``$CP$ violation from $\tau$, $t$ and $b$ dimension-6 Yukawa couplings - interplay of baryogenesis, EDM and Higgs physics,''
JHEP \textbf{05}, 056 (2020)
doi:10.1007/JHEP05(2020)056
[arXiv:2003.00099 [hep-ph]].

\bibitem{Shu:2013uua}
J.~Shu and Y.~Zhang,
Phys. Rev. Lett. \textbf{111}, no.9, 091801 (2013)
doi:10.1103/PhysRevLett.111.091801
[arXiv:1304.0773 [hep-ph]].

\bibitem{Grojean:2004xa}
C.~Grojean, G.~Servant and J.~D.~Wells,
Phys. Rev. D \textbf{71}, 036001 (2005)
doi:10.1103/PhysRevD.71.036001
[arXiv:hep-ph/0407019 [hep-ph]].

\bibitem{Chung:2012vg}
D.~J.~H.~Chung, A.~J.~Long and L.~T.~Wang,
Phys. Rev. D \textbf{87}, no.2, 023509 (2013)
doi:10.1103/PhysRevD.87.023509
[arXiv:1209.1819 [hep-ph]].

\bibitem{Gunion:1989we}
J.~F.~Gunion, H.~E.~Haber, G.~L.~Kane and S.~Dawson,
Front. Phys. \textbf{80}, 1-404 (2000)
SCIPP-89/13.

\bibitem{Djouadi:2005gj}
A.~Djouadi,
Phys. Rept. \textbf{459}, 1-241 (2008)
doi:10.1016/j.physrep.2007.10.005
[arXiv:hep-ph/0503173 [hep-ph]].

\bibitem{Branco:2011iw}
G.~C.~Branco, P.~M.~Ferreira, L.~Lavoura, M.~N.~Rebelo, M.~Sher and J.~P.~Silva,
Phys. Rept. \textbf{516}, 1-102 (2012)
doi:10.1016/j.physrep.2012.02.002
[arXiv:1106.0034 [hep-ph]].

\bibitem{DiLuzio:2020wdo}
L.~Di Luzio, M.~Giannotti, E.~Nardi and L.~Visinelli,
Phys. Rept. \textbf{870}, 1-117 (2020)
doi:10.1016/j.physrep.2020.06.002
[arXiv:2003.01100 [hep-ph]].

\bibitem{Sakurai:2021ipp}
K.~Sakurai and W.~Yin,
JHEP \textbf{04}, 113 (2022)
doi:10.1007/JHEP04(2022)113
[arXiv:2111.03653 [hep-ph]].

\bibitem{ACME:2018yjb}
V.~Andreev \textit{et al.} [ACME],
Nature \textbf{562}, no.7727, 355-360 (2018)
doi:10.1038/s41586-018-0599-8


\bibitem{LHeCStudyGroup:2012zhm}
J.~L.~Abelleira Fernandez \textit{et al.} [LHeC Study Group],
``A Large Hadron Electron Collider at CERN: Report on the Physics and Design Concepts for Machine and Detector,''
J. Phys. G \textbf{39}, 075001 (2012)
doi:10.1088/0954-3899/39/7/075001
[arXiv:1206.2913 [physics.acc-ph]].

\bibitem{LHeC:2020van}
P.~Agostini \textit{et al.} [LHeC and FCC-he Study Group],
``The Large Hadron-Electron Collider at the HL-LHC,''
J. Phys. G \textbf{48}, no.11, 110501 (2021)
doi:10.1088/1361-6471/abf3ba
[arXiv:2007.14491 [hep-ex]].


\bibitem{Han:2009pe} 
T.~Han and B.~Mellado,
``Higgs Boson Searches and the H b anti-b Coupling at the LHeC,''
Phys.\ Rev.\ D {\bf 82}, 016009 (2010)
doi:10.1103/PhysRevD.82.016009
[arXiv:0909.2460 [hep-ph]].



\bibitem{Biswal:2012mp}
S.~S.~Biswal, R.~M.~Godbole, B.~Mellado and S.~Raychaudhuri,
``Azimuthal Angle Probe of Anomalous $HWW$ Couplings at a High Energy $ep$ Collider,''
Phys. Rev. Lett. \textbf{109}, 261801 (2012)
doi:10.1103/PhysRevLett.109.261801
[arXiv:1203.6285 [hep-ph]].



\bibitem{FCC:2018byv}
A.~Abada \textit{et al.} [FCC],
``FCC Physics Opportunities: Future Circular Collider Conceptual Design Report Volume 1,''
Eur. Phys. J. C \textbf{79}, no.6, 474 (2019)
doi:10.1140/epjc/s10052-019-6904-3



\bibitem{Blumlein:1992eh} 
J.~Blumlein, G.~J.~van Oldenborgh and R.~Ruckl,
``QCD and QED corrections to Higgs boson production in charged current e p scattering,''
Nucl.\ Phys.\ B {\bf 395}, 35 (1993)
doi:10.1016/0550-3213(93)90207-6
[hep-ph/9209219].




\bibitem{Senol:2012fc} 
A.~Senol,
``Anomalous Higgs Couplings at the LHeC,''
Nucl.\ Phys.\ B {\bf 873}, 293 (2013)
doi:10.1016/j.nuclphysb.2013.04.016
[arXiv:1212.6869 [hep-ph]].



\bibitem{Hesari:2018ssq}
H.~Hesari, H.~Khanpour and M.~Mohammadi Najafabadi,
``Study of Higgs Effective Couplings at Electron-Proton Colliders,''
Phys. Rev. D \textbf{97}, no.9, 095041 (2018)
doi:10.1103/PhysRevD.97.095041
[arXiv:1805.04697 [hep-ph]].

\bibitem{Dutta:2021del}
S.~Dutta, A.~Goyal, M.~Kumar and A.~K.~Swain,
``Measuring $CP$ nature of $h\tau{\bar\tau}$ coupling at $e^-p$ collider,''
[arXiv:2109.00329 [hep-ph]].


\bibitem{Behera:2022gnr}
S.~Behera, B.~Brickwedde and M.~Schott,
``Studies on the $H\rightarrow bb$ cross-section measurement at the LHeC with a full detector simulation,''
[arXiv:2201.04037 [hep-ph]].

\bibitem{Khatibi:2014bsa}
S.~Khatibi and M.~Mohammadi Najafabadi,
Phys. Rev. D \textbf{90}, no.7, 074014 (2014)
doi:10.1103/PhysRevD.90.074014
[arXiv:1409.6553 [hep-ph]].

\bibitem{Han:2023krp}
T.~Han, S.~C.~I.~Leung and M.~Low,
[arXiv:2305.01010 [hep-ph]].

\bibitem{Ball:2012cx}
R.~D.~Ball, V.~Bertone, S.~Carrazza, C.~S.~Deans, L.~Del Debbio, S.~Forte, A.~Guffanti, N.~P.~Hartland, J.~I.~Latorre and J.~Rojo, \textit{et al.}
Nucl. Phys. B \textbf{867}, 244-289 (2013)
doi:10.1016/j.nuclphysb.2012.10.003
[arXiv:1207.1303 [hep-ph]].

\bibitem{NNPDF:2017mvq}
R.~D.~Ball \textit{et al.} [NNPDF],
Eur. Phys. J. C \textbf{77}, no.10, 663 (2017)
doi:10.1140/epjc/s10052-017-5199-5
[arXiv:1706.00428 [hep-ph]].

%
\bibitem{Alwall:2011uj}
J.~Alwall, M.~Herquet, F.~Maltoni, O.~Mattelaer and T.~Stelzer,
``MadGraph 5 : Going Beyond,''
JHEP {\bf 1106}, 128 (2011)
doi:10.1007/JHEP06(2011)128
[arXiv:1106.0522 [hep-ph]].

\bibitem{Alwall:2014bza}
J.~Alwall, C.~Duhr, B.~Fuks, O.~Mattelaer, D.~G.~Ozturk and C.~H.~Shen,
``Computing decay rates for new physics theories with FeynRules  and MadGraph 5\_aMC@NLO,''
Comput.\ Phys.\ Commun.\  {\bf 197}, 312 (2015)
doi:10.1016/j.cpc.2015.08.031
[arXiv:1402.1178 [hep-ph]].

\bibitem{Alwall:2014hca}
J.~Alwall {\it et al.},
``The automated computation of tree-level and next-to-leading order differential cross-sections, and their matching to parton shower simulations,''
JHEP {\bf 1407}, 079 (2014)
doi:10.1007/JHEP07(2014)079
[arXiv:1405.0301 [hep-ph]].

\bibitem{Degrande:2011ua} 
C.~Degrande, C.~Duhr, B.~Fuks, D.~Grellscheid, O.~Mattelaer and T.~Reiter,
``UFO - The Universal FeynRules Output,''
Comput.\ Phys.\ Commun.\  {\bf 183}, 1201 (2012)
doi:10.1016/j.cpc.2012.01.022
[arXiv:1108.2040 [hep-ph]].


\bibitem{Sjostrand:2014zea}
T.~Sjostrand {\it et al.},
``An Introduction to PYTHIA 8.2,''
Comput.\ Phys.\ Commun.\  {\bf 191}, 159 (2015)
doi:10.1016/j.cpc.2015.01.024
[arXiv:1410.3012 [hep-ph]].

\bibitem{Sjostrand:2007gs}
T.~Sjostrand, S.~Mrenna and P.~Z.~Skands,
``A Brief Introduction to PYTHIA 8.1,''
Comput.\ Phys.\ Commun.\  {\bf 178}, 852 (2008)
doi:10.1016/j.cpc.2008.01.036
[arXiv:0710.3820 [hep-ph]].

\bibitem{deFavereau:2013fsa}
J.~de Favereau {\it et al.} [DELPHES 3 Collaboration],
``DELPHES 3, A modular framework for fast simulation of a generic collider experiment,''
JHEP {\bf 1402}, 057 (2014)
doi:10.1007/JHEP02(2014)057
[arXiv:1307.6346 [hep-ex]].

\bibitem{gitFCCeh:2020}
https://github.com/delphes/delphes/blob/master/cards/delphes{\_}card{\_}FCCeh.tcl


\bibitem{gitLHeC:2020}
https://github.com/delphes/delphes/blob/master/cards/delphes{\_}card{\_}LHeC.tcl



\bibitem{Cacciari:2008gp}
M.~Cacciari, G.~P.~Salam and G.~Soyez,
``The anti-$k_t$ jet clustering algorithm,''
JHEP {\bf 0804}, 063 (2008)
doi:10.1088/1126-6708/2008/04/063
[arXiv:0802.1189 [hep-ph]].

\bibitem{Cacciari:2011ma}
M.~Cacciari, G.~P.~Salam and G.~Soyez,
``FastJet User Manual,''
Eur.\ Phys.\ J.\ C {\bf 72}, 1896 (2012)
doi:10.1140/epjc/s10052-012-1896-2
[arXiv:1111.6097 [hep-ph]].


\bibitem{Hocker:2007ht} 
A.~Hocker {\it et al.},
``TMVA - Toolkit for Multivariate Data Analysis,''
PoS ACAT {\bf }, 040 (2007)
[physics/0703039 [PHYSICS]].

\bibitem{Stelzer:2008zz} 
J.~Stelzer, A.~Hocker, P.~Speckmayer and H.~Voss,
``Current developments in TMVA: An outlook to TMVA4,''
PoS ACAT {\bf 08}, 063 (2008).

\bibitem{Speckmayer:2010zz} 
P.~Speckmayer, A.~Hocker, J.~Stelzer and H.~Voss,
``The toolkit for multivariate data analysis, TMVA 4,''
J.\ Phys.\ Conf.\ Ser.\  {\bf 219}, 032057 (2010).

\bibitem{Therhaag:2010zz} 
J.~Therhaag,
``TMVA Toolkit for multivariate data analysis in ROOT,''
PoS ICHEP {\bf 2010}, 510 (2010).

\bibitem{Xia:2018cfz}
L.~G.~Xia,
``Understanding the boosted decision tree methods with the weak-learner approximation,''
[arXiv:1811.04822 [physics.data-an]].



\bibitem{cl1}
T.~Junk,
Nucl. Instrum. Meth. A \textbf{434}, 435-443 (1999)
doi:10.1016/S0168-9002(99)00498-2
[arXiv:hep-ex/9902006 [hep-ex]].

\bibitem{cl2}
A.~L.~Read,
J. Phys. G \textbf{28}, 2693-2704 (2002)
doi:10.1088/0954-3899/28/10/313

\bibitem{Moneta:2010pm}
L.~Moneta, K.~Belasco, K.~S.~Cranmer, S.~Kreiss, A.~Lazzaro, D.~Piparo, G.~Schott, W.~Verkerke and M.~Wolf,
PoS \textbf{ACAT2010}, 057 (2010)
doi:10.22323/1.093.0057
[arXiv:1009.1003 [physics.data-an]].

\end{thebibliography}
\end{document}